\documentclass[acmtog,nonacm]{acmart}
\setcopyright{acmlicensed}
\copyrightyear{2026}
\acmYear{2026}
\acmConference[SIGGRAPH Conference Papers '26]{Special Interest Group on Computer Graphics and Interactive Techniques Conference Conference Papers}{July 19--23, 2026}{Los Angeles, CA, USA}
\acmBooktitle{Special Interest Group on Computer Graphics and Interactive Techniques Conference Conference Papers (SIGGRAPH Conference Papers '26), July 19--23, 2026, Los Angeles, CA, USA}
\acmDOI{10.1145/3799902.3811057}
\acmISBN{979-8-4007-2554-8/2026/07}

\acmSubmissionID{295}

\author{Yilin Liu}
\affiliation{
  \institution{Autodesk Research}
  \city{London}
  \country{United Kingdom}}
\authornote{Corresponding author: Yilin Liu (whatsevenlyl@gmail.com)}

\author{Pradeep Jayaraman}
\affiliation{
  \institution{Autodesk Research}
  \city{Toronto}
  \country{Canada}
}

\author{Chinthala Reddy}
\affiliation{
  \institution{Autodesk Research}
  \city{London}
  \country{United Kingdom}
}

\author{Xiang Xu}
\affiliation{
  \institution{Autodesk Research}
  \city{Toronto}
  \country{Canada}
}

\author{Hooman Shayani}
\affiliation{
  \institution{Autodesk Research}
  \city{London}
  \country{United Kingdom}
}

\usepackage{multirow}
\usepackage{xspace}
\usepackage{caption}
\usepackage{subcaption}
\usepackage{cuted}
\usepackage{float}
\usepackage{algorithm}

\newcommand{\name}{DualBrep\xspace}
\newcommand{\papertitle}{\name: A Dual-Field Continuous Representation for B-rep Modelling}

\makeatletter
\newcommand{\makesupplementtitle}{%
	\begin{strip}
		\centering
		{\@titlefont \papertitle\ Supplementary Materials\par}
	\end{strip}%
}
\makeatother

\newcommand{\rev}[1]{{#1}}

\begin{document}
\title{\papertitle}

\begin{teaserfigure}
    \includegraphics[width=\linewidth]{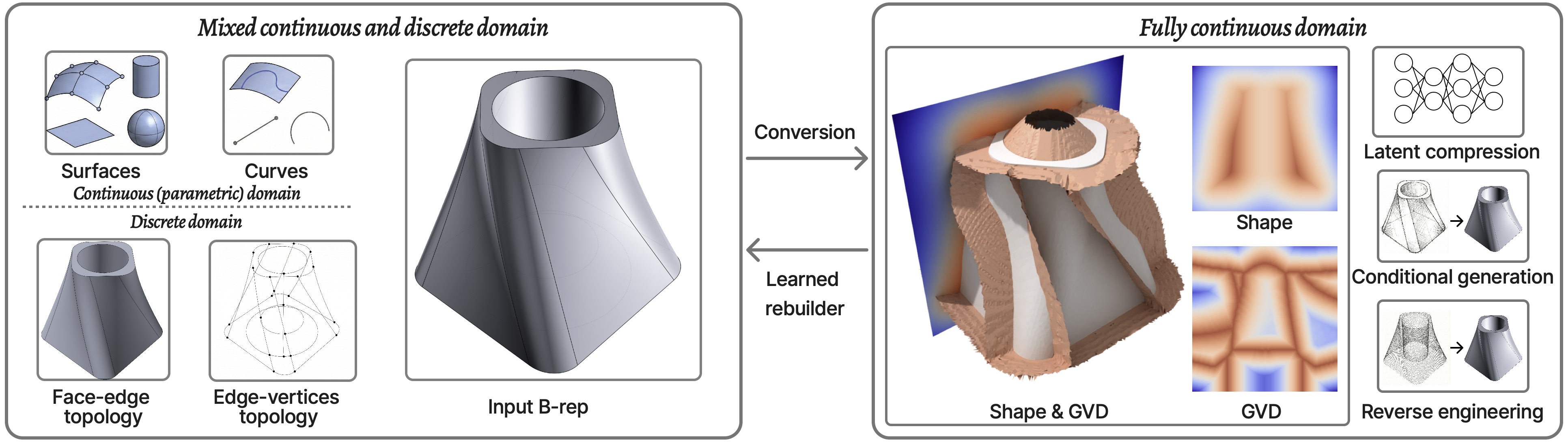}
    \caption{
        \textbf{\name bridges the discrete-continuous gap in B-rep learning.}
        Standard B-reps (left) define shapes by explicitly stitching together disjoint parametric surfaces and curves via a discrete connectivity graph, a representation that is difficult to optimize by gradient-based methods.
        \name reformulates this into a fully continuous domain (right) by encoding geometry as a Shape field and topology as a Generalized Voronoi Diagram (GVD) field.
        \rev{Both fields are compressed into a single shared latent space. This creates a backbone that supports deterministic reverse engineering through flexible UDF-guided segmentation while letting a generative model sample geometry and topology jointly.}
        A learned rebuilder extracts explicit, watertight B-rep models directly from these continuous signals.
    }
    \label{fig:teaser}
\end{teaserfigure}

\begin{abstract}
  Boundary Representation (B-rep) is the most commonly used data format in Computer-Aided Design (CAD) due to its analytical precision and direct support for parametric editing.
  However, its heterogeneous data structure— continuous parametric geometry with discrete topological graphs—poses fundamental challenges for deep learning models.
  Existing methods often \textit{directly} predict the heterogeneous B-rep graph, relying on fixed-size padding or sequential tokenization to handle the varying cardinality of the geometric primitives.
  These approaches struggle with the combinatorial complexity of CAD models. The discrete, non-differentiable nature of the graph data structure prevents end-to-end optimization of the geometry and watertightness.
  In this work we introduce \name, a novel \textit{continuous} representation that unifies B-rep geometry and topology within a fully structured Euclidean domain.
  \name encodes a CAD model using dual scalar fields: a Signed Distance Function (SDF) to represent the global shape geometry, and an Unsigned Distance Field (UDF) that implicitly encodes the topological structure via a Voronoi partitioning of the surface elements.
  \rev{Rather than processing these fields independently, we compress them into a single latent space. While the dual-field formulation alone already gives reconstruction a flexible, primitive-free segmentation signal that adapts to arbitrary face counts and surface types, the shared latent also makes generation tractable: a Flow Matching model can sample geometry and topology jointly from a single code, avoiding the error accumulation that plagues sequential or autoregressive B-rep predictors.}
  Finally, we use a neural rebuilder to extract explicit B-rep models—comprising both prismatic and free-form primitives—directly from our continuous dual scalar fields.
  We demonstrate that \name serves as a robust, unified backbone for CAD B-rep learning, achieving strong performance in both reverse engineering from raw point clouds and generative modeling via latent flow matching.
  Code is available at \url{https://github.com/AutodeskAILab/DualBrep}.
\end{abstract}




\maketitle

\section{Introduction}
Boundary Representation (B-rep) stands as the de-facto standard format in Computer-Aided Design (CAD), forming the backbone of modern manufacturing, physical simulation, and creative design pipelines.
Unlike lightweight representations such as triangle meshes or point clouds, B-reps offer analytical precision and rich editability by encoding solid shapes as a collection of parametric surfaces, trimmed by explicit topological boundaries.
With the advent of large-scale CAD datasets~\cite{ABC,willis2021fusion} and foundational geometric deep learning, there is a growing interest in modeling the distribution of B-rep data for tasks such as text-to-CAD generation~\cite{khan2024textcad,HolaBRep25,BrepGPT25}, CAD autocompletion~\cite{brepgen24,AutoBrep25}, and constrained synthesis~\cite{Casey_2025_ICCV}.
However, the inherent structure of B-reps poses a fundamental ``compatibility paradox" for deep learning.
While neural networks excel at modeling continuous signals in Euclidean space, B-rep data is intrinsically heterogeneous and combinatorial, intertwining various continuous parametric equations (geometry) with discrete graph structures (topology).
This structural mismatch makes learning a robust generative B-rep distribution exponentially difficult, especially as shape complexity increases.

Most existing approaches address this heterogeneity by modeling the discrete B-rep either as construction history~\cite{deepcad21,hnccad} or raw boundary components~\cite{solidgen23,brepgen24}. 
Even recent unified generative models~\cite{HolaBRep25,AutoBrep25,BrepGPT25} rely on discrete tokenization or fixed-size padding. 
This paradigm faces a critical limitation rooted in the complexity of B-reps: modeling the combinatorial topology distribution using gradient-based optimization creates a disjoint optimization landscape. Since the discrete topological choices are non-differentiable, the network cannot be optimized end-to-end to minimize geometric inconsistencies or to improve watertightness.
This optimization challenge is further exacerbated by the long sequences required to represent complex shapes; as the chain of dependencies grows, the generative process suffers from severe error accumulation. 
Consequently, solid watertightness diminishes rapidly~\cite{HolaBRep25,AutoBrep25}, frequently yielding broken shapes that are unusable for engineering.

In contrast, the field of mesh and surface generation has undergone a dramatic transformation driven by continuous implicit representations~\cite{park2019deepsdf,mescheder2019occupancy,wu2016learning,zhang20233dshape2vecset,trellis24,dora25,sparc3d25}. 
State-of-the-art methods like Hunyuan3D~\cite{hunyuan} and TripoSR~\cite{TripoSR24} generate highly detailed, watertight shapes with remarkable prompt fidelity.
Their success stems from a pivotal shift: rather than predicting a discrete mesh directly, they model the underlying shape as a continuous field (e.g., Signed Distance Functions (SDF) or occupancy fields). 
Since continuous fields are differentiable and resolution-independent, they are naturally aligned with gradient-based optimization, allowing networks to learn complex shape distributions without worrying about combinatorial validity.

Inspired by this dichotomy, we propose to rethink B-rep learning by shifting its representational domain from discrete graphs to continuous fields. Towards this end,
we introduce \name, a framework that models B-rep data as a fully continuous signal. 
Instead of manipulating discrete graph structures, we encode the CAD model into two spatially aligned scalar fields: i) A Signed Distance Function (SDF) that defines the global, watertight shape geometry, ensuring differentiability and resolution independence, and ii) a novel application of the Generalized Voronoi Diagram (GVD), as shown in Fig.~\ref{fig:teaser}.
Treating the B-rep faces as geometric "sites," the GVD forms a continuous sheet in 3D space that partitions the volume into regions nearest to each B-rep surface. 
We encode this structure as a \textit{single} Unsigned Distance Field (UDF) representing the distance from any point in space to this GVD boundary. 
\rev{Rather than processing these signals in isolation, we compress them into a single latent space. This unified backbone serves both reverse engineering and generative tasks under a common training pipeline. A neural rebuilder converts the decoded fields into explicit surface patches and stitches them into a watertight B-rep.}

\rev{
This continuous backbone supports both \textit{deterministic reconstruction} and \textit{generative modeling} under a single architecture, with the unified latent playing a different role in each. Most of the reconstruction quality comes from the dual-field formulation itself: encoding topology as a spatial signal lets the network infer face boundaries that adapt freely to shape complexity---without constraining surface types or requiring primitive detection---and the neural rebuilder turns the resulting segmentation into analytical patches with accurate connectivity. Sharing a latent here primarily simplifies the architecture, allowing the same backbone to be reused across both tasks rather than maintaining separate encoders for SDF and UDF. 
In generative modeling, the shared latent space is the key to consistent geometry and topology.
A Flow Matching model samples a single latent code, from which both geometry and topology are jointly decoded. This joint decoding keeps the two fields consistent with each other and avoids the error accumulation commonly seen in sequential or autoregressive B-rep prediction methods.
}

To the best of our knowledge, \name is the first framework to leverage a unified continuous domain for B-rep representation learning. 
While the final conversion back to discrete B-rep is not strictly lossless and may still introduce invalid geometry, this design allows us to defer discretization until the global geometry and topology are well established. 
Consequently, we avoid forcing the model to hallucinate topology from noise, effectively reducing the final B-rep extraction to a well-defined reconstruction task.

Through extensive experiments, we demonstrate that this paradigm effectively shifts the bottleneck of B-rep modeling. 
By leveraging our continuous dual-field backbone, \name achieves strong performance across various B-rep learning tasks, including reverse engineering from raw point clouds and conditional generation (point cloud, image) via latent flow matching models.
More importantly, we show that \name scales gracefully with shape complexity, maintaining high validity rates and low geometric and topological error as the number of B-rep primitives increases.

\section{Related Work}
\subsection{B-rep Reconstruction}
Reconstructing B-rep models from raw point clouds is a longstanding challenge in reverse engineering.
Early learning-based approaches, such as ParseNet~\cite{parsenet20}, HPNet~\cite{hpnet21}, SEDNet~\cite{sed23}, and Point2CAD~\cite{point2cad}, typically rely on segmented point sets as their core representation.
These methods treat reconstruction as a primitive fitting problem, where the point cloud is first segmented into patches, and each patch is then fitted to a parametric surface.
However, this paradigm is usually limited by the expressiveness of the input point cloud and the segmentation accuracy, often failing to capture clean boundaries or global topological consistency.

Distinct from these segmentation-based approaches, ComplexGen~\cite{complexgen22} formulates reconstruction as a detection task, using a primitive-based representation to regress surfaces, edges, and vertices directly.
The lack of explicit topological constraints often leads to models with high rates of invalid topology, such as missing or overlapping primitives.
Most relevant to our work is NVDNet~\cite{nvd24}, which introduces a structure-aware Voronoi partitioning to represent the segmentation of B-rep models.
However, NVDNet focuses on local geometric priors and only performs the reconstruction of the Voronoi diagram.
Our method extends the Voronoi diagram concept from NVDNet into a dual-distance field formulation.
\rev{The dual-field formulation itself, paired with our learned rebuilder, drives the reconstruction gains by providing a flexible, primitive-free segmentation signal and recovering free-form surfaces-a long-standing limitation of prior works. We further compress both fields into a shared latent so that the same backbone also supports conditional generation, where joint sampling of geometry and topology avoids the error accumulation typical of sequential B-rep predictors.}

\subsection{B-rep Generation}
Generative modeling of B-rep data has evolved through several paradigms.
Command-based methods, such as DeepCAD~\cite{deepcad21}, SkexGen~\cite{skexgen}, and HNC-CAD~\cite{hnccad}, represent B-reps as a sequence of construction operations (e.g., sketch and extrude).
While these methods guarantee valid CAD models by design, they are limited by the availability of construction history data and struggle to represent complex, free-form surfaces that cannot be easily described by simple operations.

Direct B-rep modeling approaches attempt to generate the B-rep data structure itself, but employ different sequential strategies.
SolidGen~\cite{solidgen23} tokenizes the B-rep into a sequence of geometric and topological elements, employing three Transformer models to autoregressively predict them.
BrepGen~\cite{brepgen24} leverages a structured latent space to represent the B-rep hierarchy, using diffusion models to generate the geometric and topological features.
DTGBrepGen~\cite{DTGBrepGen} explicitly decouples the generation of topology and geometry to reduce complexity.
However, these sequential pipelines suffer from severe error accumulation, especially when modeling complex topology graphs, frequently resulting in invalid or broken B-reps.
Recent single-stage unified predictors, including HoLa~\cite{HolaBRep25}, AutoBrep~\cite{AutoBrep25}, and BrepGPT~\cite{BrepGPT25}, aim to mitigate this by predicting the entire B-rep structure in a more holistic manner.
Nevertheless, they still face the fundamental challenge of modeling variable-length sequences to accommodate the varying cardinality of B-rep elements, which complicates the optimization landscape and fails to guarantee topological validity, often resulting in non-watertight models.

\subsection{Mesh Generation}
In contrast to the discrete struggles of B-rep generation, mesh generation has seen rapid progress driven by continuous representations.
Various continuous representations have been explored to facilitate robust modeling.
Latent-based approaches like VecSet~\cite{zhang20233dshape2vecset} and TRELLIS~\cite{trellis24} encode shapes into structured latent codes for scalable generation.
Meanwhile, methods such as CraftsMan~\cite{craftsman24}, DoRA~\cite{dora25}, Hunyuan3D~\cite{hunyuan}, and TripoSR~\cite{TripoSR24} leverage powerful neural fields or VAEs to achieve high-fidelity generation.
\rev{
These approaches build on the foundational works of DeepSDF~\cite{park2019deepsdf} and Occupancy Networks~\cite{mescheder2019occupancy}, which established learned implicit functions as a robust paradigm for 3D shape representation, and have been extended to high-resolution sparse settings by NeuralVDB~\cite{neuralvdb}.
}
Sparse voxel representations are also used by XCube~\cite{xcube24} and Sparc3D~\cite{sparc3d25} for efficient high-resolution modeling.
These representations are resolution-independent and naturally differentiable, making them exceptionally easy to model with neural networks.
Inspired by the success of these continuous approaches, we propose to bridge the gap between B-rep and mesh learning.
We introduce a dual-field representation that captures the precision of B-rep topology within a continuous Euclidean domain, allowing for robust learning and generation of complex CAD models.

\section{Method}
\label{sec:method}

\begin{figure*}[t]
    \centering
    \includegraphics[width=\linewidth]{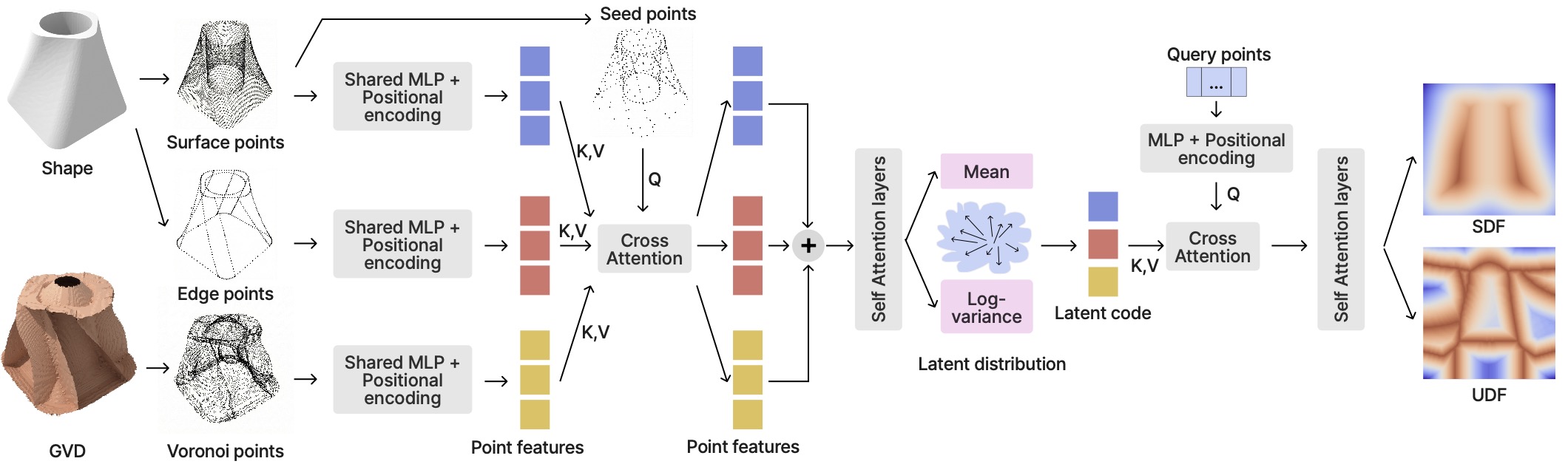}
    \caption{
        \textbf{Architecture of the VAE.} Perceiver-style encoder fuses surface, edge, and Voronoi point features into a latent representation using cross-attention. A similar decoder queries latent code to reconstruct the shape (SDF) and GVD (UDF), providing a differentiable representation of B-rep geometry and topology.
    }
    \label{fig:vae}
\end{figure*}

\subsection{Dual-Field Representation}
\label{sec:dual_field}

\paragraph{B-rep Preliminaries}
A standard Boundary Representation (B-rep) model $\mathcal{B} = (\mathcal{V}, \mathcal{E}, \mathcal{F}, \mathcal{T})$ defines a solid via a hierarchy of geometric and topological elements: vertices $\mathcal{V}$, edges $\mathcal{E}$ (parametric curves), faces $\mathcal{F}$ (parametric surfaces), and a topology graph $\mathcal{T}$.
The graph $\mathcal{T}$ encodes connectivity, specifically the face-to-edge incidences $\mathcal{T}_{f \to e}$ (defining trimming loops) and edge-to-vertex incidences $\mathcal{T}_{e \to v}$.
For the purpose of neural generation, explicitly modeling the full tuple is redundant.
Our framework focuses on recovering the minimal sufficient set: the surfaces $\mathcal{F}$, the curves $\mathcal{E}$, and the face-to-edge connectivity $\mathcal{T}_{f \to e}$.
Derivative elements, such as vertices $\mathcal{V}$ and edge-to-vertex connectivity $\mathcal{T}_{e \to v}$, can be inferred deterministically from curve intersections once the primary structure is established.

To unify the heterogeneous analytical primitives (e.g., B-splines, planes, cylinders) typically found in $\mathcal{F}$ and $\mathcal{E}$, we adopt a consistent structured discretization~\cite{brepnet23,brepgen24,uvnet21}.
Each face $f_k \in \mathcal{F}$ is represented as a regular geometry grid $\mathbf{G}_k \in \mathbb{R}^{M \times M \times 3}$, generated by uniformly sampling the surface within its parametric bounds $[u_{\min}, u_{\max}] \times [v_{\min}, v_{\max}]$.
\rev{For periodic surfaces (e.g., full cylinders or tori), we cut the surface along the seam line to obtain a single open parameterization and overlap the last column or row of the grid with the first, so that the resulting grid has the same $M \times M$ layout while preserving continuity at the seam.}
Similarly, each edge $e_j \in \mathcal{E}$ is sampled into a linear grid $\mathbf{C}_j \in \mathbb{R}^{M \times 3}$, where $M$ is the number of sample points (set to 16 in all our experiments).

\paragraph{Continuous Dual-Field Formulation}
As shown in Fig.~\ref{fig:teaser}, to circumvent the optimization difficulties inherent to discrete topology graphs, we map the B-rep entirely into the continuous Euclidean domain.
Our key insight is that a B-rep can be fundamentally viewed as a watertight geometric hull that is partitioned into distinct regions (faces).
We capture this duality by encoding the model into two complementary scalar fields—one representing the geometry and the other representing the structural boundaries (topology).

\paragraph{Geometry field}
We use a standard Signed Distance Function $\mathcal{S}: \mathbb{R}^3 \to \mathbb{R}$ to capture the global shape.
Its zero-level set, $\mathcal{S}(\mathbf{p}) = 0$, defines the continuous, watertight hull of the object, ignoring the internal segmentation of faces.

\paragraph{Topology Field (GVD)}
This is the critical component that recovers the B-rep structure.
In a B-rep, topological edges exist exactly where two distinct surfaces meet.
We capture this relationship volumetrically using the Generalized Voronoi Diagram (GVD).
Conceptually, the GVD partitions the ambient 3D space into "Voronoi cells," where every point in a cell is uniquely closest to a single B-rep face $f_i$.
The boundaries between these cells form continuous "medial sheets" in 3D space.
We encode these sheets as an Unsigned Distance Field (UDF) $\mathcal{U}: \mathbb{R}^3 \to \mathbb{R}$, where $\mathcal{U}(\mathbf{p})$ gives the distance from point $\mathbf{p}$ to the nearest point on the GVD surface.

\paragraph{Why this representation}
The power of this formulation lies in the intersection of these fields.
The SDF defines \textit{where the surface exists}, while the GVD defines \textit{where the surface identity changes}.
\begin{itemize}
    \item $\mathcal{S}(\mathbf{p}) \approx 0$: The point lies on the object surface.
    \item $\mathcal{U}(\mathbf{p}) \approx 0$: The point is equidistant to multiple faces (i.e., it lies on a medial sheet or edge).
\end{itemize}
Therefore, the superposition of $\mathcal{S}$ and $\mathcal{U}$ provides a complete definition of the B-rep: the SDF recovers the geometry, and the GVD implicitly "cuts" this geometry into the correct topological patches.
This allows us to learn complex topologies without needing to predict discrete adjacency matrices or handle the variable cardinality of faces, as the GVD naturally adapts to any number of partitions.

\subsection{Dual-Field Variational Autoencoder} \label{sec:vae}
We employ a Variational Autoencoder (VAE) to compress this dual-field representation into a single, compact latent space. Inspired by recent neural field architectures~\cite{zhang20233dshape2vecset,dora25}, we design a Perceiver-style encoder that uses cross-attention to aggregate multi-modal inputs, as illustrated in Fig.~\ref{fig:vae}.

\paragraph{Input Representation}
To capture high-frequency details, we sample three point sets as input:
(i) Surface points $\mathcal{P}_s$ from the shape geometry ($N_s = 32,768$);
(ii) Edge points $\mathcal{P}_e$ along B-rep curves ($N_e = 32,768$); and
(iii) Voronoi points $\mathcal{P}_v$ from the GVD sheets ($N_v = 32,768$).
These sets collectively describe the surface, the sharp features, and the topological boundaries.

\begin{figure*}[t]
    \centering
    \includegraphics[width=\linewidth]{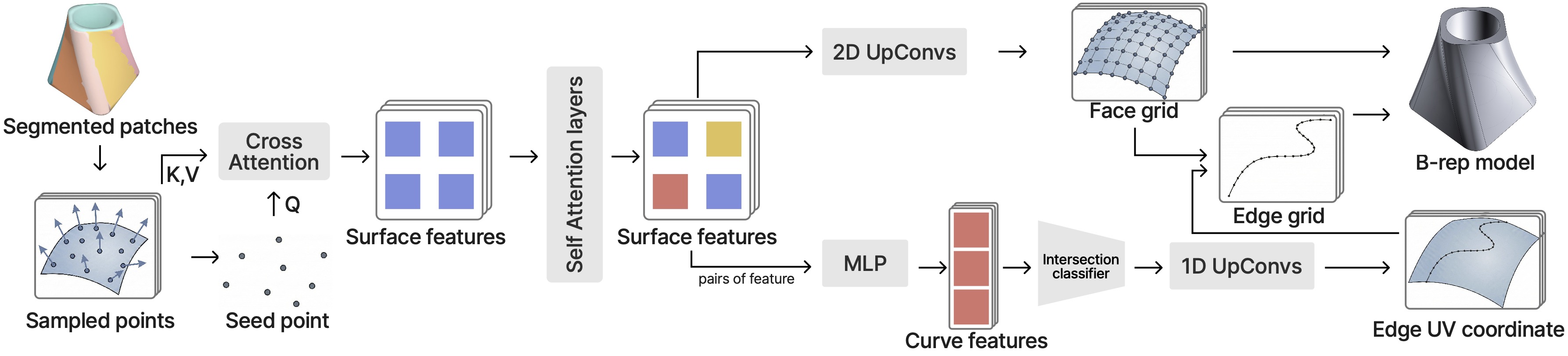}
    \caption{
        \textbf{Architecture of the learned rebuilder.} Segmented face patches are encoded into features and processed through self-attention to capture topological context. The network then predicts patch-level parametric UV grids, adjacency relationships, and UV-space trimming curves, which are assembled into a B-rep.
    }
    \label{fig:rebuilder}
\end{figure*}

\paragraph{Encoder Architecture}
The encoder maps these inputs to a latent embedding $\mathbf{Z} \in \mathbb{R}^{K \times D}$, with sequence length $K=2048$ and dimension $D=32$ (split into mean and log-variance).
We first embed the spatial coordinates of each set $\mathcal{P}_m$ ($m \in \{s, e, v\}$) using a shared MLP with a shared sinusoidal positional encoding $\gamma(\cdot)$:
\begin{equation}
    \mathbf{F}_m = { \text{MLP}_{\text{in}}(\gamma(\mathbf{p})) \mid \mathbf{p} \in \mathcal{P}_m }.
\end{equation}
To form the latent bottleneck, we select $K$ anchor points from $\mathcal{P}_s$ via Farthest Point Sampling to serve as queries $\mathbf{Q}$.
To disentangle the contributions of geometry and topology, the queries attend to each input modality \textit{individually} via cross-attention:
\begin{equation}
    \mathbf{H}_m = \text{CrossAttn}(\mathbf{Q}, \mathbf{F}_m), \quad m \in \{s, e, v\}.
\end{equation}
These modality-specific features are fused via a weighted sum and refined via self-attention layers to enable global information exchange, before being projected to the variational parameters $\boldsymbol{\mu}, \log \boldsymbol{\sigma}^2$ as the latent distribution $Z \sim \mathcal{N}(\boldsymbol{\mu}, \boldsymbol{\sigma})$.

\paragraph{Decoder Architecture}
The decoder acts as an implicit neural function.
Given a query coordinate $\mathbf{x} \in \mathbb{R}^3$, we extract local context by attending to the sampled latent codes $\mathbf{z} \sim \mathcal{N}(\boldsymbol{\mu}, \boldsymbol{\sigma})$:
\begin{equation}
    \mathbf{h}_x = \text{CrossAttn}(\text{MLP}_q(\gamma(\mathbf{x})), \mathbf{z}).
\end{equation}
After a set of self-attention layers, two separate MLP heads then regress the field values: $\hat{s}(\mathbf{x}) = \text{MLP}_\text{SDF}(\mathbf{h}_x)$ and $\hat{u}(\mathbf{x}) = \text{MLP}_\text{UDF}(\mathbf{h}_x)$.

\paragraph{Training Objectives}
We train the VAE end-to-end using a hybrid sampling strategy for query points $\mathbf{x}$, drawing from the bounding volume, near-surface regions, and near-edge regions in a 1:1:2 ratio~\cite{dora25}.
The loss function is the sum of $L_1$ reconstruction loss and KL regularization terms with a weight of $0.001$.
The VAE has $\approx$250M parameters, consisting of 16 and 8 self-attention layers in the encoder and decoder respectively.
Each layer has 8 attention heads and a hidden dimension of 1024.

\paragraph{Deterministic Mode for Reverse Engineering}
For the task of reverse engineering (reconstructing B-reps from raw point clouds), we adapt the VAE into a \textit{deterministic} autoencoder and train the model from scratch.
We mask out the edge and Voronoi inputs ($\mathcal{P}_e, \mathcal{P}_v$) and set the weight of KL loss to 0.
This forces the encoder to infer the complete dual-field structure—including topological segmentation—solely from the surface geometry $\mathcal{P}_s$.

\subsection{Latent Flow Matching for Generation} \label{sec:generation}

To enable scalable and controllable generation, we train a conditional Flow Matching~\cite{flowmaching2023} model over the latent space $\mathcal{Z}$.

\paragraph{Unified Geometric Conditioning}
While our framework is agnostic to the conditioning modality $\mathbf{c}$, we prioritize \textit{point clouds} as a primary modality due to their widespread availability and adoption.
We encode point cloud conditions using a Perceiver-style encoder similar to our VAE, aggregating features from $N_c = 1024$ seed points.
Motivated by its potential for future editing applications, we also include results for native single-view image generation, where the conditional signal is obtained by a pre-trained DINOv2~\cite{oquab2023dinov2} ViT.
Other modalities (e.g., text embeddings from LLMs) can be similarly integrated.

\paragraph{Architecture and Inference}
We implement the flow matching model using a Diffusion Transformer (DiT)~\cite{Peebles2022DiT} adapted for set-structured latents (16 layers, 1024 dim, $\approx$300M parameters).
Conditioning is injected via cross-attention, and time $t$ modulates features via Adaptive Layer Norm (AdaLN).
During inference, we solve the ODE using the Euler method (50 steps) to obtain $\hat{\mathbf{z}}$.
This code is decoded into dual fields $\mathcal{S}$ and $\mathcal{U}$.

\rev{
\paragraph{Mesh Extraction and Segmentation}
We extract the surface mesh via Marching Cubes on $\mathcal{S}$ and segment it into a segmented mesh $\mathcal{M}_\text{seg}$ via hierarchical region growing constrained by the boundaries in $\mathcal{U}$. More concretely, we evaluate the UDF on the Marching-Cubes mesh and use it as a boundary cue on the face-adjacency graph. A coarse pass extracts provisional surface patches, and a second stricter pass revisits ambiguous components to split merged faces when necessary. The final labels define the segmented mesh $\mathcal{M}_\text{seg}$ passed to the rebuilder; additional implementation details are provided in the supplementary material.
}

\subsection{Learned B-rep Rebuilder}\label{sec:rebuilder}
The final component is a learned rebuilder that converts the explicit segmented mesh $\mathcal{M}_\text{seg}$ into an analytical B-rep.
Unlike heuristic fitting methods~\cite{nvd24}, our neural rebuilder directly predicts parametric structures, offering robustness to noise and support for free-form surfaces.

\paragraph{Problem Formulation}
Given $N$ disjoint face patches $\{\mathcal{P}_1, \ldots, \mathcal{P}_N\}$ from the segmented mesh $\mathcal{M}_\text{seg}$, the rebuilder predicts:
(i) \textit{Surface Geometry:} A structured grid $\mathbf{G}_k$ for each patch;
(ii) \textit{Topology:} An adjacency matrix $\mathbf{A} \in \{0, 1\}^{N \times N}$;
(iii) \textit{Trim Curves:} For each connected pair $(i, j)$, a parametric curve $\mathbf{C}_{ij}$ defining the shared boundary.

\paragraph{Feature Encoding}
We represent each patch $\mathcal{P}_k$ by sampling $N_p=100$ points (positions and normals).
A lightweight Perceiver-style encoder maps these points to a patch feature vector $\mathbf{h}_k^{(0)} \in \mathbb{R}^{4 \times 256}$, where $4$ is the number of subsampled seed points per patch and $256$ is the feature dimension.
To capture topological context, these features are processed by a global self-attention module:
\begin{equation}
    [\mathbf{h}_1, \ldots, \mathbf{h}_N] = \text{SelfAttn}([\mathbf{h}_1^{(0)}, \ldots, \mathbf{h}_N^{(0)}]),
\end{equation}
where $\mathbf{h}_k \in \mathbb{R}^{4 \times 256}$ is the refined feature for patch $\mathcal{P}_k$.

\paragraph{Surface Geometry Head}
This head reconstructs the parametric surface geometry.
For each patch, we first reshape the learned feature tokens $\mathbf{h}_k \in \mathbb{R}^{4 \times 256}$ into a spatial feature map of size $2 \times 2 \times 256$.
This feature map is progressively upsampled via a sequence of 2D transposed convolution blocks (ConvTranspose $\to$ ReLU $\to$ BatchNorm) to resolutions $4^2$, $8^2$, and finally $16^2$:
\begin{equation}
    \mathbf{G}_k = \text{ConvTranspose}_{2\text{D}}(\text{Reshape}(\mathbf{h}_k, [2, 2, 256])) \in \mathbb{R}^{16 \times 16 \times 3}.
\end{equation}
The output grid $\mathbf{G}_k$ represents 3D control points sampled uniformly in the patch's UV parameter domain, which are subsequently used to fit the final B-spline surface.

\paragraph{Edge and Trim Prediction}
To determine connectivity and boundaries between two patches $i$ and $j$, we first construct an edge feature by concatenating their token sequences: $\mathbf{e}_{ij} = [\mathbf{h}_i; \mathbf{h}_j] \in \mathbb{R}^{8 \times 256}$.
This feature is flattened to a vector of size $2048$ and processed by an MLP to predict the adjacency probability $\hat{a}_{ij}$.
For active edges ($\hat{a}_{ij} > 0.5$), we regress the trim curve.
Unlike prior works~\cite{brepgen24,HolaBRep25} that predict curves in 3D Euclidean space, we predict the curve in the 2D UV domain of its corresponding surface to ensure consistency with the surface parameterization.
The flattened feature is projected and reshaped into a 1D sequence, then progressively upsampled via 1D transposed convolutions to produce the final curve $\mathbf{C}_{ij} \in \mathbb{R}^{16 \times 2}$:
\begin{equation}
    \mathbf{C}_{ij} = \text{ConvTranspose}_{1\text{D}}(\text{MLP}_{\text{proj}}(\text{Flatten}(\mathbf{e}_{ij}))).
\end{equation}

\paragraph{Canonicalization Strategy}
A key challenge is the ambiguity of UV mappings (e.g., arbitrary rotation). We enforce a \textit{canonical pose} during training: for every ground-truth surface, we sort the sampled points lexicographically (ZYX) and define the point with the minimum value as the UV origin. This ensures the network learns a pose-invariant parameterization.

\paragraph{B-rep Rebuilding}
Similar to previous works~\cite{brepgen24,HolaBRep25}, the final B-rep is assembled deterministically:
(i) B-spline surfaces are fitted to the predicted grids $\mathbf{G}_k$;
(ii) Predicted 2D trim points $\mathbf{C}_{ij}$ are projected onto these surfaces to form 3D edges;
(iii) The topology graph is built from the predicted adjacency matrix $\hat{\mathbf{A}}$, and the model is stitched using a standard CAD kernel.
\rev{Note that the current rebuilder fits all surfaces as B-splines without predicting analytic surface or curve types (e.g., plane, cylinder, circle). Incorporating type classification to recover exact primitives where applicable is left to future work.}

\begin{table*}[t]
    \centering
    \caption{Quantitative comparison of point cloud to B-rep reconstruction on ABC dataset.
        We report the number of predicted primitives versus GT primitives, Chamfer Distance (CD, lower is better) for geometric accuracy, F1-score (\%) for primitive detection accuracy and topological correctness, and overall validity rate (\%).
        Best results are in \textbf{bold}, second best are \underline{underlined}.
    }
    \label{tab:main_results}
    \resizebox{\textwidth}{!}{
        \begin{tabular}{l|ccc|ccc|ccc|cc|c}
            \toprule
            \multirow{2}{*}{Method}   & \multicolumn{3}{c|}{Primitive Count (\#)} & \multicolumn{3}{c|}{Geometric Accuracy (CD $\downarrow$)} & \multicolumn{3}{c|}{Primitive F1-score (\%) $\uparrow$} & \multicolumn{2}{c|}{Topology F1-score (\%) $\uparrow$} & Validity                                                                                                                                                                     \\
                                      & Surface                                   & Edge                                                      & Vertex                                                  & Surface                                                & Edge               & Vertex             & Surface             & Edge                & Vertex              & Face-Edge           & Edge-Vertex         & Rate (\%) $\uparrow$ \\
            \midrule
            \rev{SEDNet+Point2CAD$^\dagger$}          & 7.9/19.2                                  & 18.2/48                                                   & 12.7/31.5                                               & 0.0259                                                 & 0.0336             & 0.1499             & 48.68\%             & 45.01\%             & 41.21\%             & 40.87\%             & 33.65\%             & /                    \\
            \rev{NVDNet$^\dagger$}                    & 22.8/19.2                                 & 67.7/48                                                   & 53.2/31.5                                               & \textbf{0.0142}                                        & \textbf{0.0059}    & \underline{0.0301} & \textbf{83.70\%}    & 78.48\%             & 71.92\%             & \textbf{80.84\%}    & 74.27\%             & \rev{12\%$^\ddagger$}                    \\
            HoLa-BRep                  & \underline{16.0/19.2}                     & \textbf{44.5/48}                                          & \textbf{29.7/31.5}                                      & 0.0211                                                 & 0.0329             & 0.0750             & 79.33\%             & 71.65\%             & 65.08\%             & 68.81\%             & 66.41\%             & \underline{73.98\%}  \\
            \midrule
            Ours\textsubscript{gen}   & \textbf{19.3/19.2}                        & 55.8/48                                                   & 40.2/31.5                                               & 0.0157                                                 & 0.0133             & 0.0247             & 77.18\%             & \underline{86.68\%} & \underline{81.33\%} & 70.31\%             & \underline{84.91\%} & 69.49\%              \\
            Ours\textsubscript{recon} & \textbf{19.3/19.2}                        & \underline{53.9/48}                                       & \underline{37.6/31.5}                                   & \underline{0.0156}                                     & \underline{0.0129} & \textbf{0.0203}    & \underline{81.98\%} & \textbf{89.87\%}    & \textbf{84.74\%}    & \underline{76.36\%} & \textbf{88.17\%}    & \textbf{76.34\%}     \\
            \bottomrule
        \end{tabular}
    }
\end{table*}

\begin{figure}[t]
    \centering
    \begin{subfigure}[b]{1\linewidth}
         \centering
         \includegraphics[width=\linewidth]{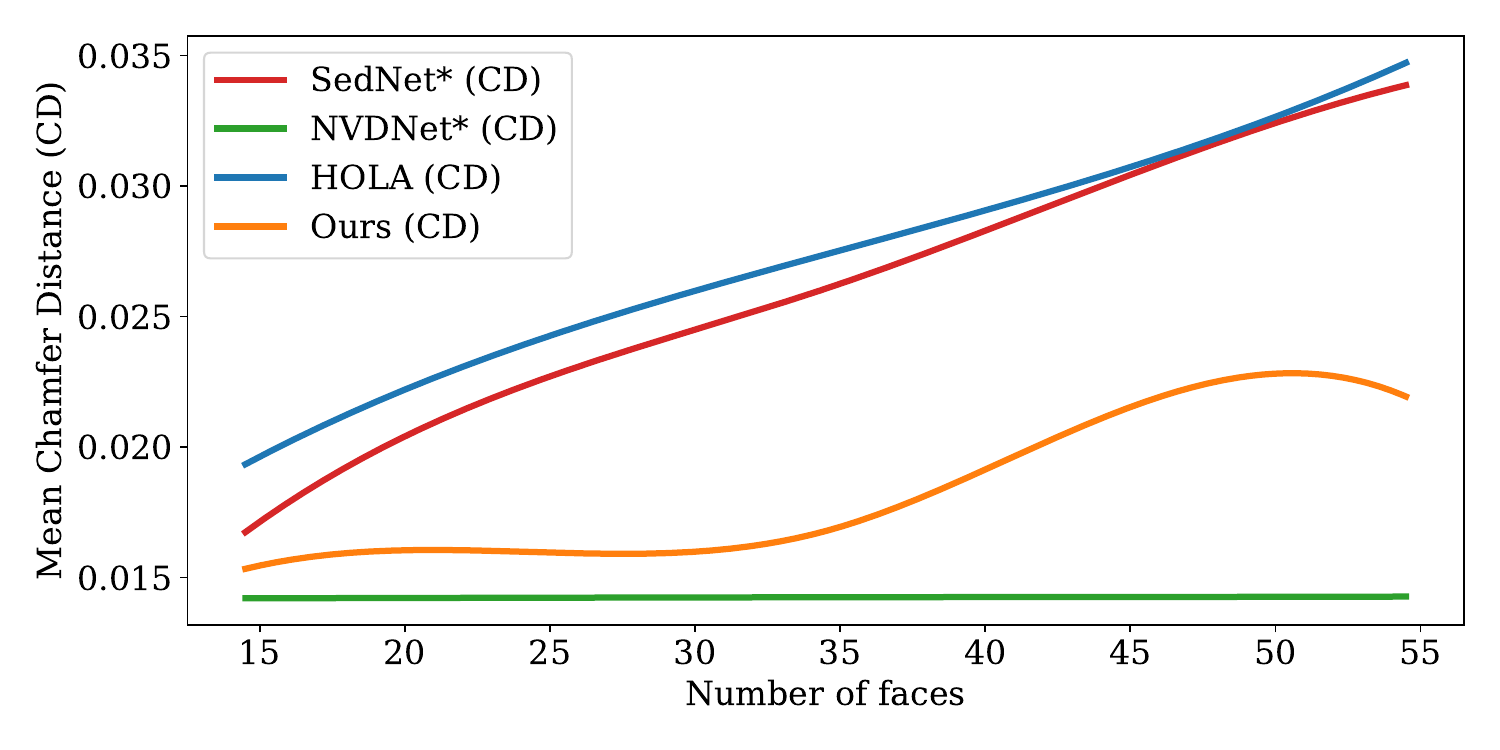}
         \caption{Geometric error vs. shape complexity}
        \label{fig:chamfer_vs_complexity}
     \end{subfigure}
    \hfill
    \begin{subfigure}[b]{1\linewidth}
         \centering
         \includegraphics[width=\linewidth]{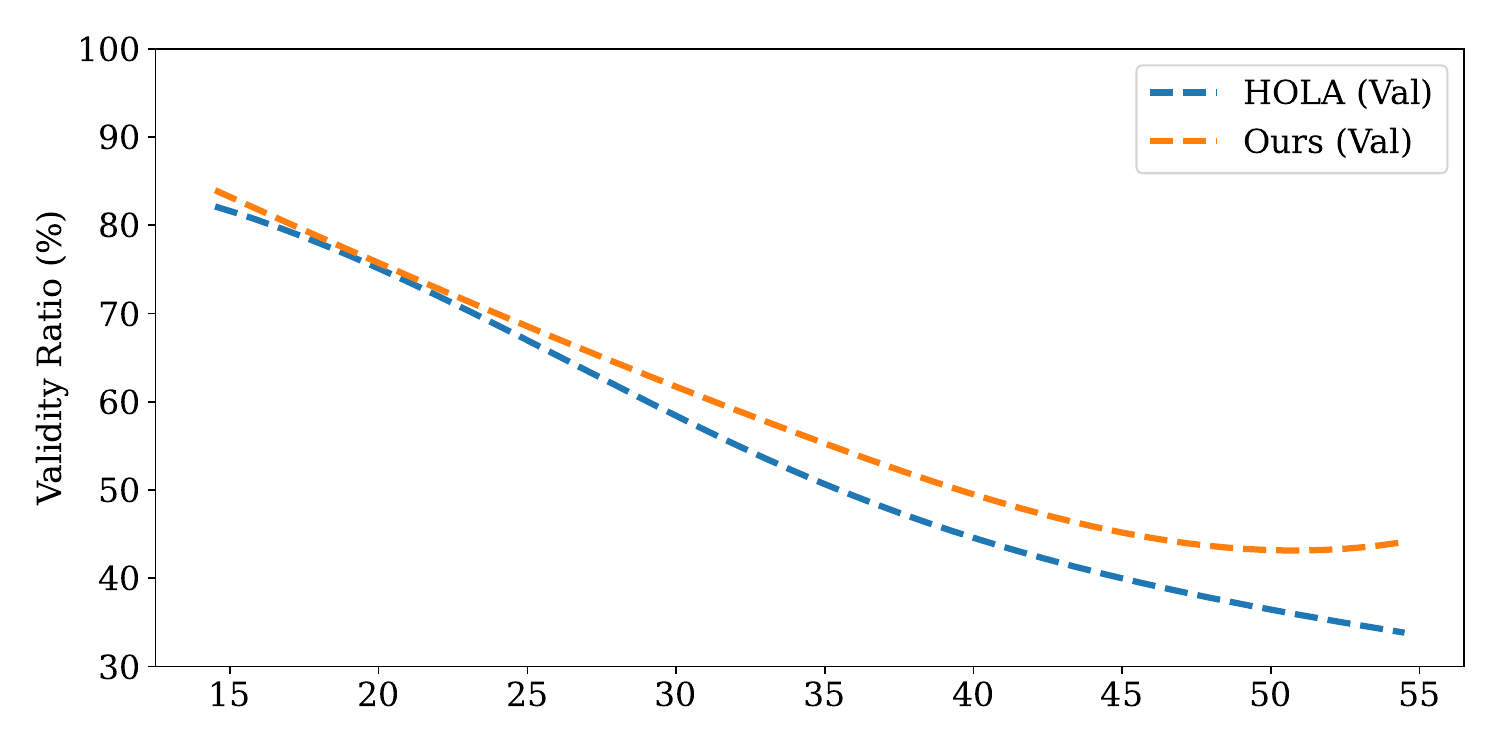}
         \caption{Validity vs. shape complexity}
        \label{fig:validity_vs_complexity}
     \end{subfigure}
    \caption{
        \textbf{Reconstruction performance vs. shape complexity.} We analyze how the reconstruction performance of different methods varies with shape complexity, measured by the number of faces in the B-rep model. \textit{Top}: Chamfer Distance (lower is better) vs. number of faces. \textit{Bottom}: Validity rate (higher is better) vs. number of faces. Our \name framework maintains stable performance across different shape complexities, outperforming baseline methods in both geometric accuracy and validity, especially for complex shapes with many faces. Note that curves are interpolating B-splines of degree 3 fit to the data.
    }
    \label{fig:complexity_analysis}
\end{figure}

\section{Results}
We validate \name primarily as a robust backbone for deterministic reverse engineering—converting point clouds into watertight, editable CAD models. Additionally, we demonstrate that our unified latent representation constructs a high-quality manifold for generative tasks, enabling multi-modal generation via flow matching.

\subsection{Experimental Setup}
To evaluate the proposed \name framework, we treat it as the backbone model for point-cloud-to-B-rep reverse engineering tasks.
We conduct experiments on the ABC dataset~\cite{ABC}, which contains a large collection of CAD models with corresponding B-rep representations.
To include more diverse and complex CAD models, we use the complete ABC dataset instead of the DeepCAD~\cite{deepcad21} subset, filtering out models that are too simple (i.e., consisting of fewer than 10 faces) or too complex (consisting of more than 100 faces or multiple solids), as well as duplicated samples. \rev{Detailed filtering criteria are provided in the supplementary material.}
After filtering, we obtain a dataset of $\sim$80k CAD models, where 4k models are used for testing and the rest are for training and validation.

\paragraph{Metrics}
Following prior works~\cite{nvd24,HolaBRep25}, we evaluate the performance of our method using the following metrics:
\begin{itemize}
    \item \textit{Geometric accuracy:} We use Chamfer Distance to measure the geometric accuracy of the reconstructed B-rep surface, edges and vertices against their ground truth counterparts.
    \item \textit{Primitive accuracy:} In addition to geometric accuracy that measures the global shape similarity, we also evaluate how the parametric primitives are grouped to form the final B-rep structure. We measure the detection-based scores including precision, recall, and F1-score for each primitive type (surfaces, edges, vertices).
    \item \textit{Topological accuracy:} We evaluate the correctness of the predicted B-rep topology by comparing the predicted and ground truth adjacency matrices of B-rep elements (faces, edges, vertices). We report the accuracy, precision, recall, and F1-score for each adjacency matrix.
    \item \textit{Validity:} We also report the percentage of valid B-rep models among all reconstructed models, where a valid B-rep model is defined as one that satisfies the topological and geometric constraints of a well-formed B-rep~\cite{brepgen24}.
\end{itemize}

\rev{\footnotetext{$\dagger$ Trained on the smaller ParseNet split of ABC.}}
\rev{\footnotetext{$\ddagger$ NVDNet's mesh decomposition post-processed with our B-rep parameterization and CAD-kernel assembly for a comparable validity evaluation.}}

\paragraph{Baselines}
We compare our \name framework with several state-of-the-art methods for point-cloud-to-B-rep reconstruction, including segmentation-based methods (SEDNet+Point2CAD~\cite{point2cad,sed23}), a Voronoi-based method (NVDNet~\cite{nvd24}), and generative methods that take point clouds as input conditions~\cite{HolaBRep25}.
\rev{See supplementary for details on baseline implementations and training.}

\subsection{Reverse Engineering}
Table~\ref{tab:main_results} summarizes the performance on the point-cloud-to-B-rep reconstruction task. \name outperforms baseline methods across the majority of metrics, achieving a notable validity ratio of 76.34\%. This result supports our main claim: the dual-field representation, together with UDF-guided segmentation and the learned rebuilder, forms a strong continuous backbone for B-rep recovery.

While NVDNet achieves a marginally lower Chamfer Distance, it optimizes for raw point-to-surface distance without guaranteeing structural integrity. 
Lacking explicit UV parameterization, NVDNet is restricted to fitting standard prismatic primitives and relies on alpha-shapes for trimming—a heuristic that frequently yields non-watertight models incompatible with CAD software. In contrast, \name explicitly models UV grids for both faces and edges. This allows us to recover intricate edge loops and vertex relationships with superior accuracy. Consequently, although our raw Chamfer distance is slightly higher, our framework delivers models that are topologically valid and engineering-ready, rather than just geometrically close clouds of surfaces.
\rev{To quantify this, we applied our B-rep parameterization and rebuilding to NVDNet's mesh decomposition (denoted NVDNet$^\ddagger$ in Table~\ref{tab:main_results}). Without a global prior, NVDNet's local cues produce severe over-segmentation and disjointed boundaries; its validity drops to 12\% and its surface CD rises to 0.0367 (vs.\ our 0.0156 at 76.34\% validity), confirming its initial geometric advantage largely vanishes once structural integrity is enforced.}

\begin{figure*}[t]
    \centering
    \includegraphics[width=1\linewidth]{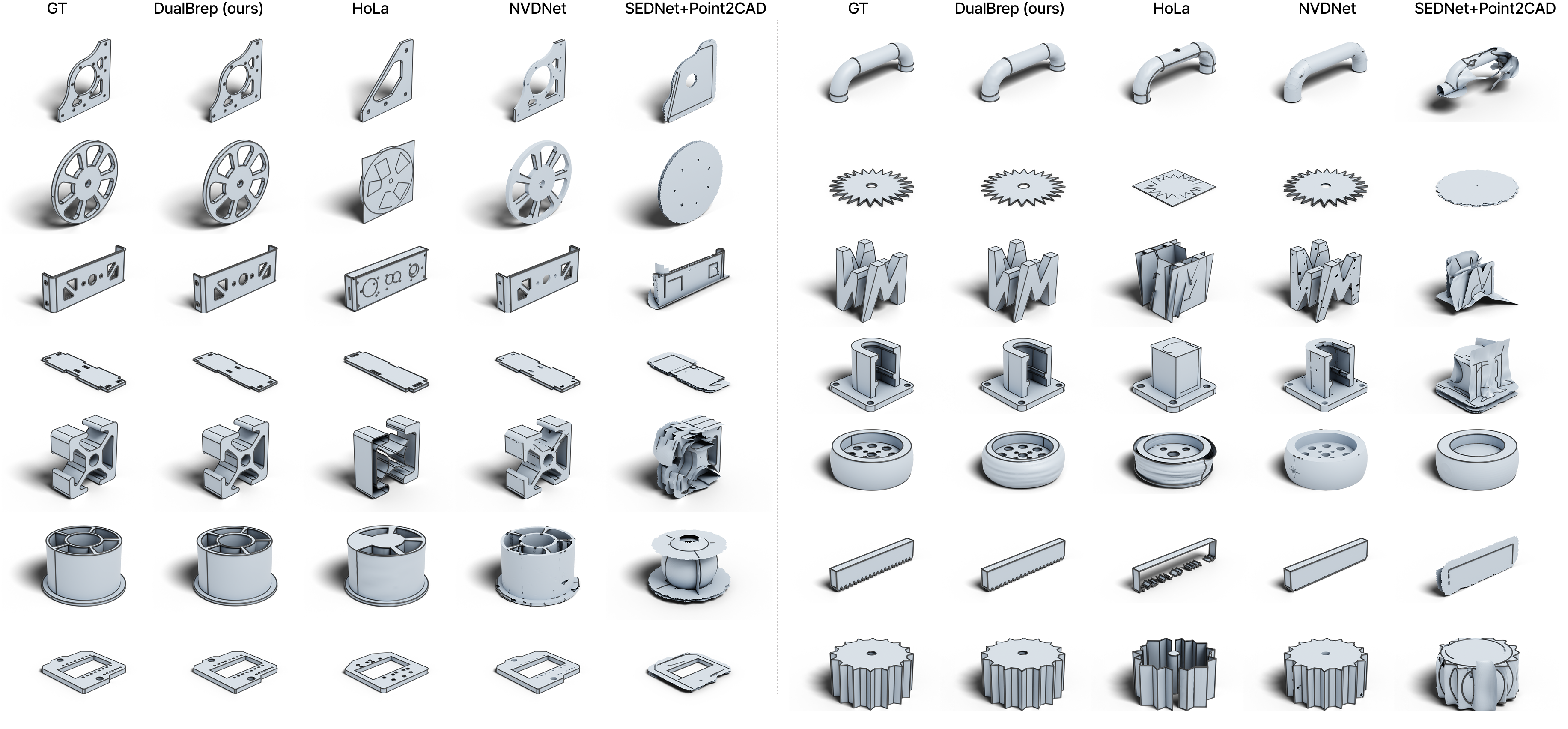}
    \caption{
        \textbf{Qualitative comparison on point-cloud-to-B-rep reconstruction.} 
        We compare our \name{}\textsubscript{recon} with baseline methods on various CAD models. From left to right: ground truth B-rep, \name (Ours), HoLa, NVDNet, and SEDNet+Point2CAD. 
        Our method produces more accurate surface segmentation and better preserves geometric details while maintaining topological validity, even on complex shapes like gears with 20+ teeth or mechanical parts with numerous holes and cutouts.
    }
    \label{fig:qualitative_results}
\end{figure*}

We investigated how performance degrades as shape complexity increases (Fig.~\ref{fig:complexity_analysis}). Discrete methods like HoLa struggle due to the combinatorial explosion of the topological search space. 
Predicting discrete adjacency graphs becomes exponentially difficult as the number of faces grows, making it hard to ensure global validity. 
Conversely, Voronoi-based methods like NVDNet and \name maintain stable performance regardless of face count due to their continuous field representation. 
However, \name distinguishes itself by combining this scalability with global consistency. 
Unlike NVDNet, which relies solely on local cues and often produces disjointed elements, our global dual-field backbone ensures structural coherence for both B-rep faces and edges. 
As shown in Fig.~\ref{fig:qualitative_results}, our method faithfully reconstructs intricate details—such as free-form surfaces, gears with 20+ teeth, and thin-walled frames—where baselines often fail.

Finally, we report results for the generative variant of \name trained via latent flow matching. 
Interestingly, this model yields slightly lower reconstruction metrics than the deterministic version. 
\rev{We attribute this to the stochastic nature of the ODE solver, which introduces small latent-code perturbations that, while geometrically imperceptible, are amplified by the sensitivity of the B-rep rebuilder (see supplementary). Although HoLa-BRep slightly outperforms \name{}\textsubscript{gen} on validity and surface F1, \name{}\textsubscript{gen} still dominates every edge- and vertex-level metric since predicting trim curves in UV space (vs.\ HoLa's free 3D regression) forces edges to lie strictly on the surface; even restricted to their failure cases, our edge CD/F1 is $0.0270$/$0.67$ vs.\ HoLa's $0.0508$/$0.55$.}
Nevertheless, its competitive performance confirms the robustness of our latent space, demonstrating its potential for broader CAD generation tasks beyond direct reconstruction.

\begin{figure}[t]
    \centering
    \includegraphics[width=\columnwidth]{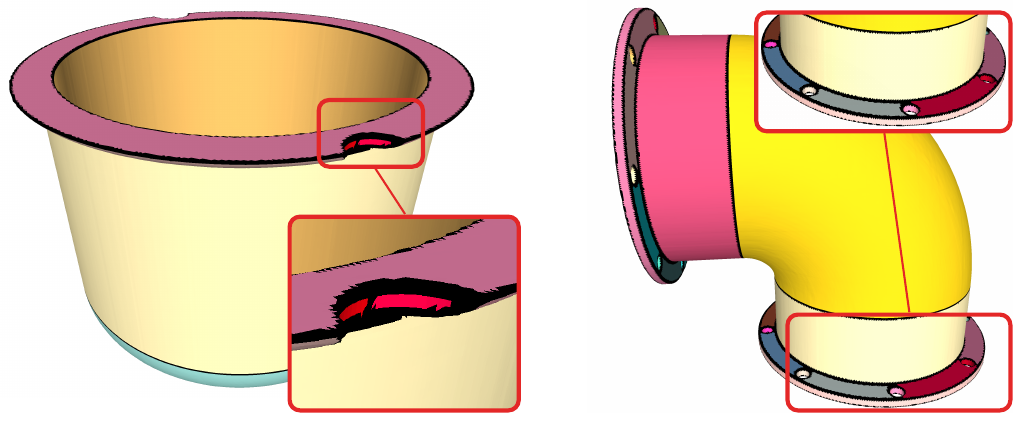}
    \caption{
        \textbf{Failure cases and limitations.}
        While we showed that cases with narrow or thin structures can be handled well by our \name framework, extremely thin features may still be lost during the segmentation, leading to invalid B-rep models.
    }
    \label{fig:failure_cases}
\end{figure}

\begin{figure*}[t]
    \centering
    \includegraphics[width=\linewidth]{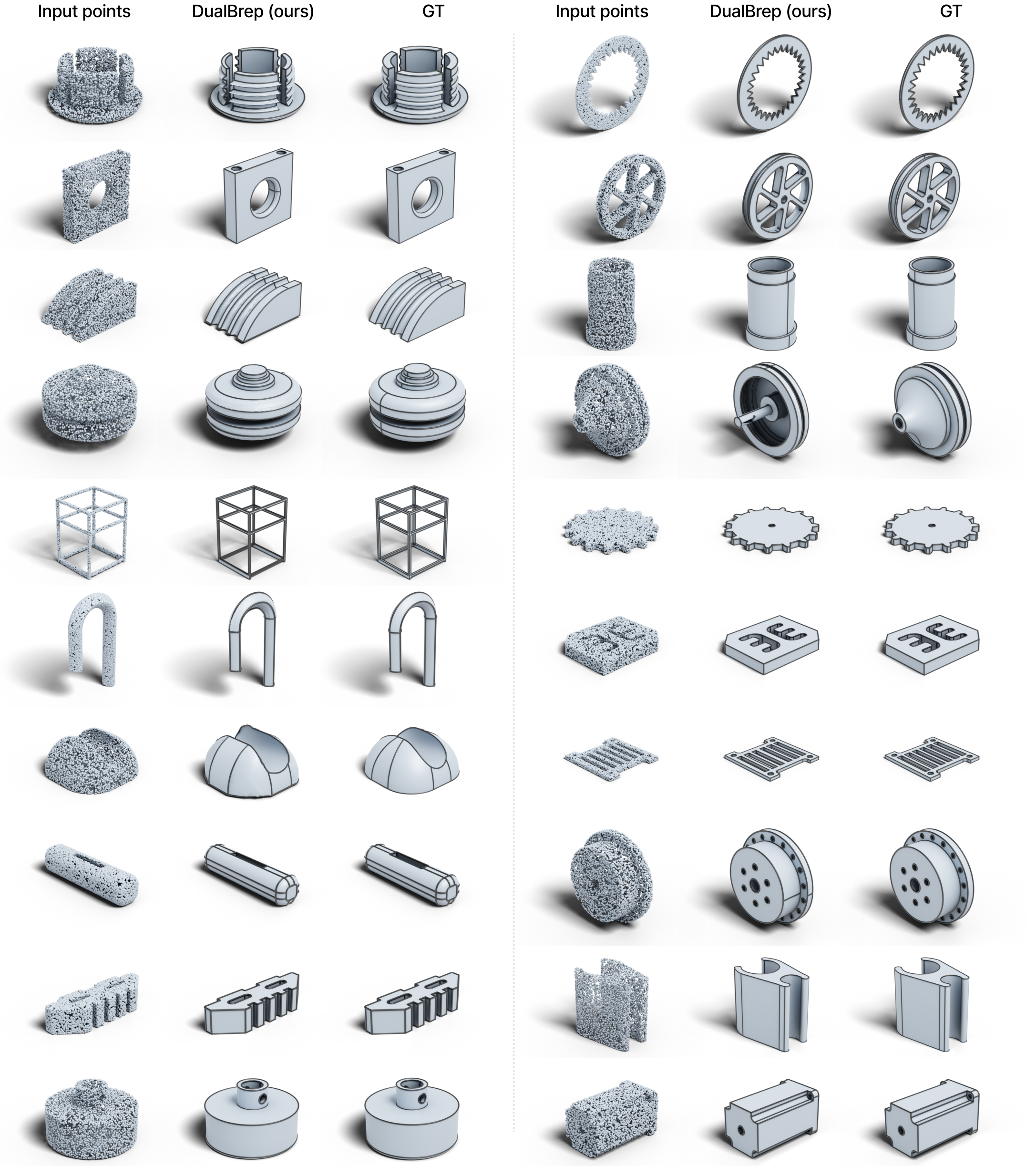}
    \caption{
        \textbf{Point-cloud-to-B-rep reconstruction gallery.} We showcase diverse reconstruction results from our deterministic \name{}\textsubscript{recon} across various CAD model categories, including shapes with free-form surfaces, mechanical parts with intricate details and thin-walled structures.
    }
    \label{fig:reconstruction_gallery}
\end{figure*}

\subsection{Native Image-to-B-rep Generation}
\begin{figure*}[t]
    \centering
    \includegraphics[width=\linewidth]{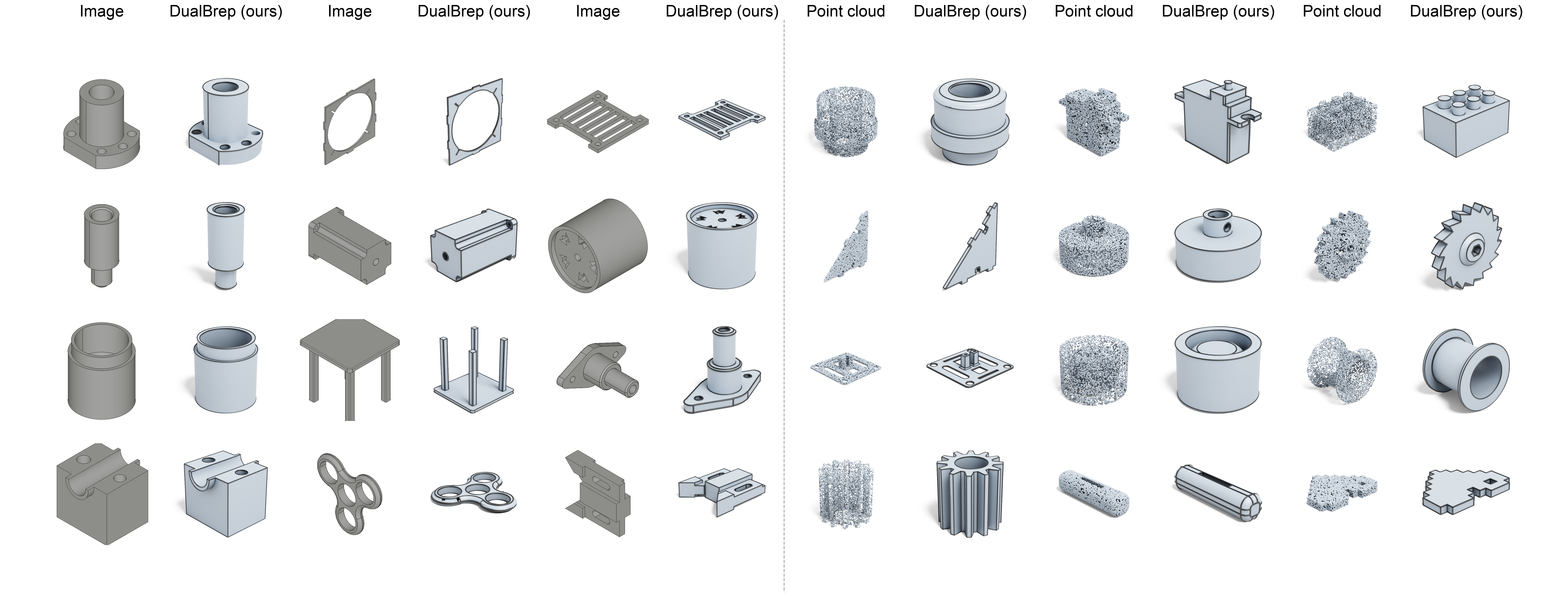}
    \caption{
        \textbf{Native conditional generation.} 
        Single view image or point cloud can be injected as conditions to our latent flow matching model \name{}\textsubscript{gen} for direct B-rep generation.
        We show various results of point-cloud-to-B-rep generation (right) and image-to-B-rep generation (left).
    }
    \label{fig:native_generation}
\end{figure*}

We further demonstrate the versatility of our latent representation by training a Latent Flow Matching model for image-conditioned generation.
As shown in Fig.~\ref{fig:native_generation}, the model generates watertight B-rep assemblies directly from single-view RGB images.
Unlike deterministic reconstruction, this generative approach learns the conditional distribution, imagining plausible details in occluded regions while respecting the global structure of the input, and maintains high topological validity and geometric fidelity even on free-form curvature and intricate mechanical details.

\subsection{Limitations and Failure Cases}
Several limitations remain. First, as a volumetric approach, our method is bounded by the grid resolution of the latent fields: extremely thin features or high-frequency details below the sampling rate may be aliased or lost. As illustrated in Fig.~\ref{fig:failure_cases}, thin-walled structures can vanish or merge with adjacent regions during GVD segmentation, leading to validity failures.
Second, the final B-rep extraction relies on a neural rebuilder, which in rare cases involving complex intersections may fail to stitch patches perfectly, producing minor non-watertight artifacts. Future work could mitigate these issues via adaptive octree sampling or hybrid explicit-implicit representations~\cite{sparc3d25}, decoupling fine-scale feature preservation from global grid resolution.

\section{Conclusion}
We introduced \name, a continuous dual-field representation that couples B-rep geometry and topology in a shared latent space. By decoding this latent into an SDF for shape and a GVD-derived UDF for topological segmentation, the framework defers discretization until global structure is established, supporting both reverse engineering and conditional generation while sidestepping the optimization difficulties of discrete graph prediction.

More importantly, this representation opens a direction beyond B-rep generation itself. Being continuous and differentiable, it can be coupled with physics-based objectives such as finite element analysis, computational fluid dynamics, and stress simulation, enabling joint optimization of geometry, manufacturability, and structural performance—a path toward CAD systems that reason about form, function, and performance together.

\bibliographystyle{ACM-Reference-Format}
\bibliography{main.bib}

\clearpage

\setcounter{section}{0}
\setcounter{table}{0}
\setcounter{figure}{0}
\makesupplementtitle
This supplementary material provides additional implementation details and extended results that complement the main paper.

\section{Construction of the Generalized Voronoi Diagram}

\rev{We construct the GVD as an explicit triangle mesh and derive the UDF from point-to-mesh distance queries. The full procedure is summarized in Algorithm~\ref{alg:voronoi}. Given a watertight B-rep with face set $\mathcal{F}$ and edge set $\mathcal{E}$, we use the following implementation.}

\paragraph{Normalization and meshing}
\rev{
We first load the STEP solid, center it at the origin, and apply a uniform scale so that the longest box dimension fits inside $[-0.9,0.9]^3$ while preserving aspect ratio.
The normalized B-rep is triangulated by OpenCascade with linear deflection $10^{-1}$. This mesh is used only for sampling and area estimation; all projections and normal evaluations are performed on the analytic surfaces and curves.

\paragraph{Point Sampling}
We build a labeled point set whose labels correspond to B-rep faces.
We first enumerate all faces and collect their non-degenerate edges.
Seam edges are removed by detecting duplicated oriented edges on the same face, and curves whose two incident sides belong to the same face are discarded.
After this filtering, every retained edge is required to be shared by exactly two distinct faces.

For each remaining shared edge $e_{ij}$ between faces $f_i$ and $f_j$, we sample points along the analytic edge with density $1000$ points per unit length, with at least $10$ samples.
To avoid numerical issues during the Delaunay triangulation, the first and last samples are moved approximately $10^{-3}$ away from the two edge endpoints, and each parameter sample receives a small random jitter of magnitude at most $10^{-6}$.
At each edge sample $\mathbf{p}$, we compute the edge tangent $\mathbf{t}$ and the two surface normals from the adjacent analytic surfaces.
For each adjacent face, we form an in-surface direction orthogonal to the edge by
\[
\mathbf{b}=\mathbf{n}\times \mathbf{t},
\]
followed by orientation correction using the face and edge orientations.
We then translate $\mathbf{p}$ by a small offset $\epsilon_v=10^{-4}$ along $\mathbf{b}$ and keep the point only if it is classified as inside the corresponding trimmed face.
This creates two nearby labeled samples, one for each side of the shared edge, which induce the Voronoi sheet separating $f_i$ and $f_j$.

We complement these edge-adjacent samples with interior surface samples.
For each face, we Poisson-sample its triangulated mesh using radius $0.005$.
Samples whose distance to any retained boundary curve on that face is smaller than $5\times 10^{-3}$ are rejected so that interior samples do not overwrite the narrow two-sided samples inserted near true inter-face boundaries.
All retained surface points inherit the label of their source face.

Finally, we add six grids of unlabeled auxiliary points on the faces of a box at coordinate $\pm 2$ (with small jitter).
These points are assigned label $-1$ and are used only to bound otherwise unbounded Voronoi cells and stabilize the extraction near the exterior domain.

\paragraph{Voronoi sheet extraction}
Rather than directly triangulating the sampled points themselves, we build a 3D Delaunay triangulation over the full labeled point set and recover the Voronoi cell of each site.
Each Voronoi cell is triangulated into polygonal facets using Geogram's convex-cell routine.
The GVD is then defined as the subset of Voronoi facets that are shared by two neighboring cells with different non-negative face labels.
For every site we compare each triangulated Voronoi facet against the facets of neighboring cells returned by the Delaunay data structure; a facet is kept only when the same triangle appears in both cells up to numerical tolerance ($10^{-12}$ squared distance).
Facets adjacent to unlabeled auxiliary points or to sites with the same face label are discarded.
The union of all retained triangles forms the GVD mesh $\mathcal{M}_{\mathrm{GVD}}$.

\paragraph{UDF Computation}
During dataset generation, the mesh $\mathcal{M}_{\mathrm{GVD}}$ is treated as a triangle soup.
For any query point $\mathbf{x}$, we compute
\[
\mathcal{U}(\mathbf{x}) = \min_{\mathbf{y}\in \mathcal{M}_{\mathrm{GVD}}} \lVert \mathbf{x}-\mathbf{y} \rVert_2,
\]
using a BVH closest-triangle query.
This produces a continuous unsigned field whose zero-set coincides with the extracted Voronoi sheets and whose low-value region localizes face-transition boundaries used later for segmentation.
}

\rev{\rev{
\begin{algorithm}[t]
    \centering
    \begin{minipage}{0.98\columnwidth}
    \small
    \hrule
    \vspace{0.6ex}
    \textbf{Algorithm 1.} Voronoi-sheet construction from a watertight B-rep.

    \textbf{Input:} watertight B-rep solid $\mathcal{B}$ with analytic faces and trimmed edges.\\
    \textbf{Output:} triangle mesh $\mathcal{M}_{\mathrm{GVD}}$ approximating the generalized Voronoi diagram between face regions.

    \begin{enumerate}
        \item Normalize $\mathcal{B}$ by centering it and uniformly scaling it to fit inside $[-0.9,0.9]^3$.
        \item Triangulate each face with OpenCascade using deflection $10^{-1}$; use this mesh only for sampling and area estimation.
        \item Enumerate all faces and build the edge-to-face map; discard degenerate edges, seam duplicates, and edges whose two incident sides belong to the same face.
        \item Initialize an empty labeled point set $\mathcal{P}$.
        \item For each retained shared edge $e_{ij}$ between faces $(f_i,f_j)$, sample at least $10$ edge parameters with density $1000$ points per unit length; move the two endpoint samples by about $10^{-3}$ away from the vertices and add a small random jitter.
        \item For each sampled edge point $\mathbf{p}$ and for each incident face $f\in\{f_i,f_j\}$, evaluate the edge tangent $\mathbf{t}$ and surface normal $\mathbf{n}$, compute $\mathbf{b}=\mathbf{n}\times\mathbf{t}$ with orientation correction, set $\mathbf{q}=\mathbf{p}+\epsilon_v\mathbf{b}$ with $\epsilon_v=10^{-4}$, and insert $\mathbf{q}$ into $\mathcal{P}$ with label $f$ if $\mathbf{q}$ remains inside the trimmed face.
        \item For each face $f_k$, Poisson-sample its triangulated surface using radius $0.005$, reject any sample within $5\times10^{-3}$ of a retained boundary curve, and insert the remaining interior samples into $\mathcal{P}$ with label $k$.
        \item Add unlabeled auxiliary samples (label $-1$) on the six faces of a box at coordinate $\pm 2$ to close unbounded exterior cells.
        \item Deduplicate $\mathcal{P}$ and remove samples too close to retained B-rep vertices.
        \item Build a 3D periodic Delaunay triangulation on $\mathcal{P}$ and extract the Voronoi cell of every site.
        \item For each site with non-negative face label, triangulate its Voronoi cell and compare each facet against the facets of Delaunay-neighbor cells having a different non-negative label.
        \item Retain a facet only if the same triangle appears in both cells within numerical tolerance; append all retained facets to $\mathcal{M}_{\mathrm{GVD}}$.
        \item Merge duplicate vertices and polygons in $\mathcal{M}_{\mathrm{GVD}}$, then export the resulting triangle mesh.
    \end{enumerate}
    \vspace{0.4ex}
    \hrule
    \end{minipage}
    \caption{Pseudo-code for the implemented Voronoi-sheet construction.}
    \label{alg:voronoi}
\end{algorithm}}
}

\section{Hierarchical Region Growing for Surface Segmentation}
\rev{

After decoding the dual fields, we first extract a triangle mesh from the SDF using Marching Cubes and then segment this mesh into face patches using the decoded UDF.
Our implementation uses a two-pass hierarchical procedure on the face-adjacency graph of the reconstructed mesh rather than an explicit queue-based flood fill.
The procedure is equivalent to region growing under UDF-defined masks, but is more concise to implement as connected-components extraction on the induced adjacency subgraphs.

Let $u_i = |\mathcal{U}(\mathbf{c}_i)|$ denote the absolute UDF value evaluated at the centroid $\mathbf{c}_i$ of triangle $i$.
In the first pass, we keep only triangles with $u_i \geq \tau_1$, extract connected components on the restricted face-adjacency graph, and discard small or weak components.
In the second pass, we revisit only the surviving components whose maximum UDF value exceeds a higher threshold $\tau_2$, and attempt to split them again using the stricter mask $u_i \geq \tau_2$.
The original component is replaced only if this second pass yields at least two valid sub-components; otherwise it is preserved.
In our implementation, we use $\tau_1=0.005$, $\tau_2=0.01$, a minimum component size of $5$, and a minimum average-UDF threshold of $0.003$. See Algorithm~\ref{alg:region_growing} for the full pseudo-code.
}

\rev{\rev{
\begin{algorithm}[t]
    \centering
    \begin{minipage}{0.98\columnwidth}
    \small
    \hrule
    \vspace{0.6ex}
    \textbf{Algorithm 2.} Hierarchical region growing on the reconstructed surface mesh.

    \textbf{Input:} Marching-Cubes mesh $\mathcal{M}$, per-face unsigned distances $u_i = |\mathcal{U}(\mathbf{c}_i)|$, thresholds $\tau_1 < \tau_2$, minimum component size $s_{\min}$, and minimum average-UDF threshold $\bar{u}_{\min}$.\\
    \textbf{Output:} face labels defining segmented surface patches.

    \begin{enumerate}
        \item Construct the face-adjacency graph $G=(V,E)$ of $\mathcal{M}$, where each node corresponds to one triangle.
        \item Initialize all labels to $-1$.
        \item \textbf{Pass 1: coarse region growing.}
        \begin{enumerate}
            \item Mark a face $i$ as active if $u_i \geq \tau_1$.
            \item Restrict $G$ to edges whose two incident faces are both active.
            \item Extract connected components on this restricted graph and assign one provisional label to each component.
            \item For each provisional component, compute its size and its average UDF value.
            \item Discard any provisional component whose size is smaller than $s_{\min}$ or whose average UDF is not larger than $\bar{u}_{\min}$.
        \end{enumerate}
        \item For every surviving pass-1 component, compute its maximum UDF value.
        \item Mark a surviving component as a split candidate if its maximum UDF value is at least $\tau_2$.
        \item \textbf{Pass 2: hierarchical re-splitting.}
        \begin{enumerate}
            \item Mark a face $i$ as eligible if it belongs to a split candidate and satisfies $u_i \geq \tau_2$.
            \item Restrict the original face-adjacency graph to edges whose two incident faces are both eligible and currently belong to the same pass-1 component.
            \item Extract connected components on this stricter graph to obtain candidate sub-components.
            \item Discard any candidate sub-component whose size is smaller than $s_{\min}$.
            \item For each split candidate component:
            \begin{enumerate}
                \item If fewer than two valid sub-components remain, keep the original pass-1 component unchanged.
                \item Otherwise, remove the original pass-1 label and replace it with one new label per valid sub-component.
            \end{enumerate}
        \end{enumerate}
        \item Relabel all non-negative labels so that they become consecutive integers.
        \item Return the final face labels.
    \end{enumerate}
    \vspace{0.4ex}
    \hrule
    \end{minipage}
    \caption{Pseudo-code for the hierarchical region-growing procedure used to segment the reconstructed surface mesh from the decoded UDF. The algorithm is implemented through connected-components extraction on thresholded face-adjacency graphs, which is equivalent to region growing under the corresponding masks.}
    \label{alg:region_growing}
\end{algorithm}}
}

\section{Dataset Filtering Criteria}
\rev{Starting from the complete ABC dataset, we apply the following filters to obtain our training and evaluation set:
\begin{itemize}
    \item \textit{Face count:} We discard models with fewer than 10 or more than 100 B-rep faces, as well as models containing multiple disjoint solids.
    \item \textit{Edge complexity:} Models containing any face with more than 50 trimming edges are removed, as such faces typically arise from degenerate or pathological geometry.
    \item \textit{Tiny primitives:} After normalizing each model to fit within $[-0.9,0.9]^3$, we discard models that contain any edge whose arc length is less than $10^{-3}$, since such features fall below the resolution of our volumetric sampling.
    \item \textit{Duplicates:} We remove near-duplicate models identified by matching face/edge counts and low chamfer distances between sampled point clouds.
\end{itemize}
After filtering, approximately 80k models remain. We randomly hold out 4k models for testing; the rest are used for training and validation. All baselines are evaluated on the same 4k test set.}

\rev{
\section{Implementation Details of Baselines}
For all baselines we use the released code and checkpoints. HoLa-BRep is evaluated with its checkpoint trained on the full ABC dataset under the same face-count filtering as ours. NVDNet and SEDNet+Point2CAD only release checkpoints trained on the smaller ParseNet~\cite{parsenet20} split of ABC ($\sim$30k models).
}

\begin{figure*}[t]
    \centering
    \includegraphics[width=1\linewidth]{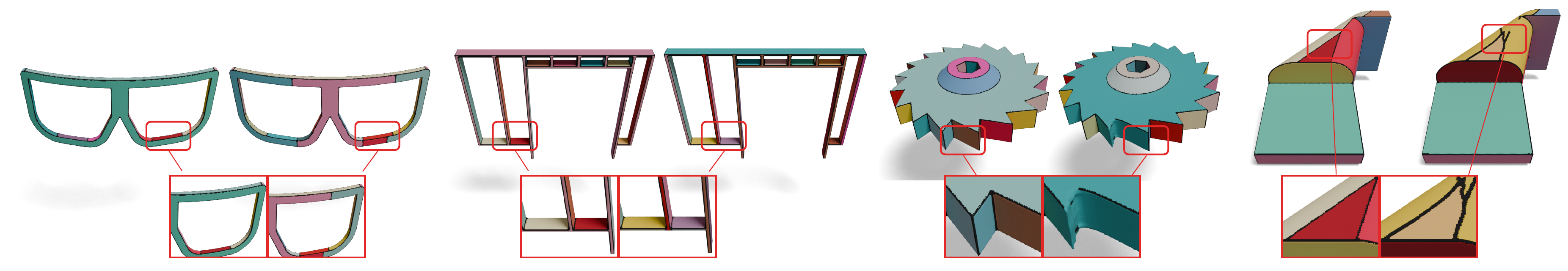}
    \caption{
        \rev{\textbf{Deterministic vs.\ generative output on the same input.} The two variants produce nearly identical geometry and segmentation. Differences are localized to one or two patches (highlighted), yet sufficient to break watertightness.}
    }
    \label{fig:validity_gap_examples}
\end{figure*}

\section{Training Details}

All models are trained using AdamW optimizer with learning rate $10^{-4}$ and default $\beta$ parameters.

\paragraph{VAE} Trained with batch size 16 on 16 H100 GPUs for approximately 4 days until convergence. We use KL weight 0.001 and hybrid spatial sampling (33\% uniform, 33\% near-surface, 33\% near-GVD boundaries).

\paragraph{Flow Matching Model} Trained with batch size 6 on 32 H100 GPUs for approximately 3 days. Uses 50-step Euler ODE solver for inference.

\paragraph{Rebuilder} Trained with batch size 16 on 8 H100 GPUs for approximately 7 days. Multi-task loss weights: $\lambda_s=1$, $\lambda_a=0.001$, $\lambda_c=1$ for surface, adjacency, and curve losses respectively.

\begin{figure}[t]
    \centering
    \includegraphics[width=\columnwidth]{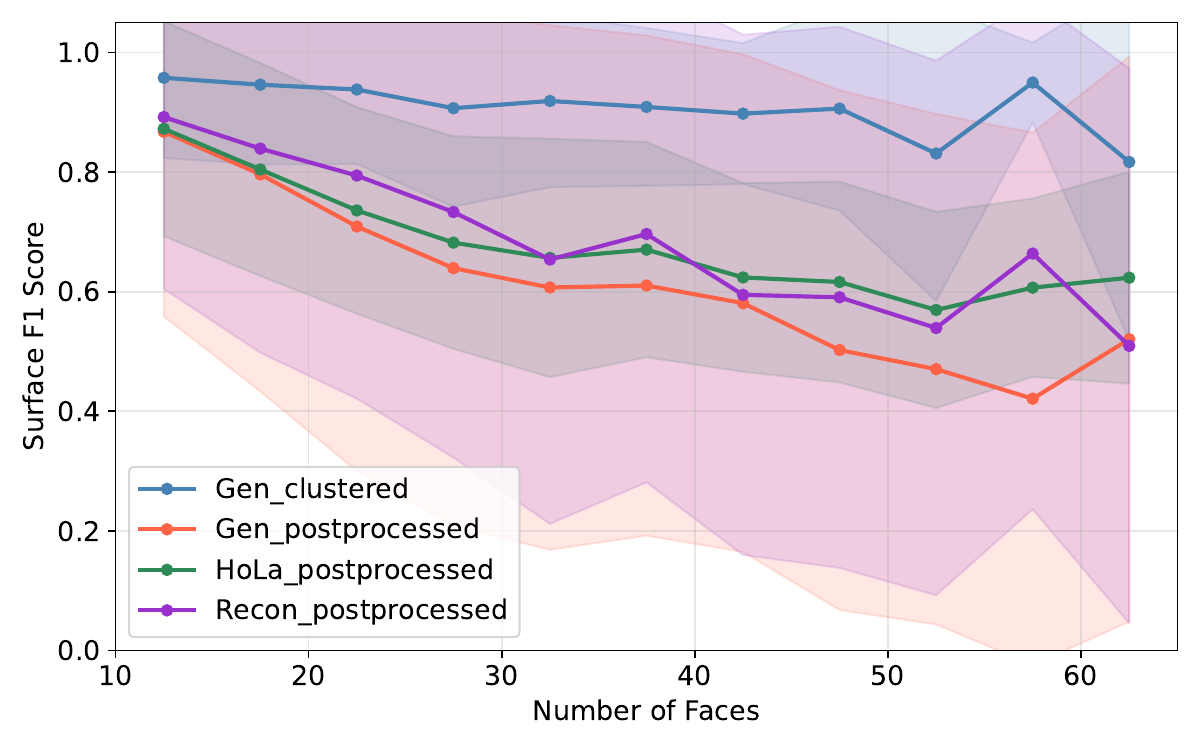}
    \caption{
    \rev{
        \textbf{Surface F1-score vs.\ shape complexity at different pipeline stages.} \name{}\textsubscript{gen}~\textit{clustered} (after segmentation only) consistently outperforms all post-rebuilding variants, confirming that patch-level continuous representations are easier to produce accurately. The wider gap for \name{}\textsubscript{gen} between \textit{clustered} and \textit{processed} indicates that the stochastic sampling introduces small perturbations that are amplified by the fragile rebuilder.
    }}
    \label{fig:f1_vs_faces}
\end{figure}

\begin{figure}[t]
    \centering
    \includegraphics[width=\columnwidth]{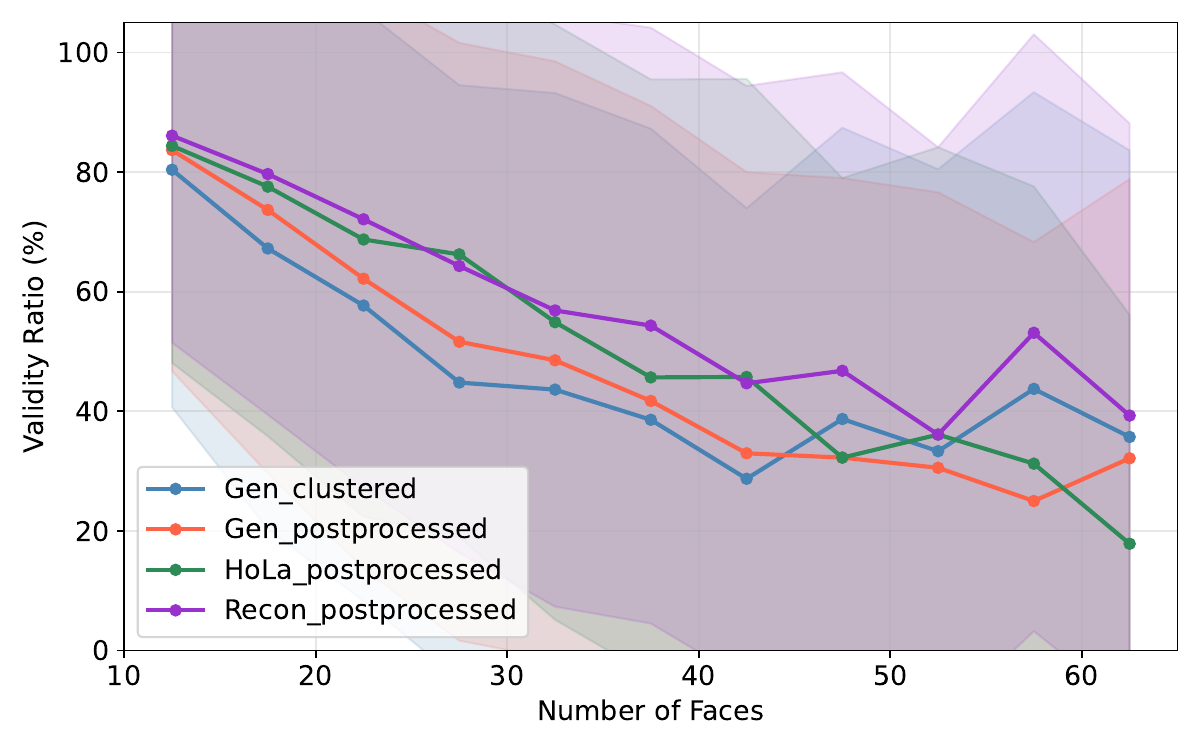}
    \caption{
    \rev{
        \textbf{Validity rate vs.\ shape complexity.} Despite strong segmentation F1, validity decreases sharply for \name{}\textsubscript{gen} after B-rep rebuilding, illustrating that even small local errors can break the watertight CAD constraint.
    }}
    \label{fig:validity_vs_faces}
\end{figure}

\rev{
\section{Understanding the B-rep Modeling Bottleneck}
\label{sec:validity_analysis}

As discussed in Sec.~4.2 of the main paper, the final conversion from continuous fields to discrete B-rep relies on a neural rebuilder that must produce a perfectly closed shell.
Here we analyze how this last-mile rebuilding step affects the relationship between primitive-level accuracy and model-level validity.

\paragraph{Primitive accuracy vs.\ solid validity.}
As shown in Table 1 in the main paper, \name{}\textsubscript{gen} achieves competitive or superior geometric accuracy (e.g., edge and vertex CD) compared to HoLa-BRep, yet its validity rate is lower.
To understand this gap, Fig.~\ref{fig:f1_vs_faces} visualizes the surface F1-score at different pipeline stages as a function of shape complexity.
We plot \name{}\textsubscript{gen} after segmentation only (\textit{clustered}) and after full B-rep rebuilding (\textit{processed}), together with the processed outputs of \name{}\textsubscript{recon} and HoLa-BRep for reference.
Among all variants, \name{}\textsubscript{gen}~\textit{clustered} achieves the highest F1 across all face counts, consistent with the finding in NVDNet~\cite{nvd24} that triangle-soup--based patch representations are inherently easier to generate and control than fully structured B-rep outputs.
However, once the rebuilder assembles these patches into a closed B-rep, the F1 of \name{}\textsubscript{gen}~\textit{processed} drops considerably, and its validity degrades even more sharply (Fig.~\ref{fig:validity_vs_faces}).
To further quantify this, we partition the \name{}\textsubscript{gen}~\textit{clustered} results into \emph{perfect} samples (surface F1\,=\,1, i.e., every ground-truth face is correctly segmented) and the rest: as shown in Fig.~\ref{fig:validity_vs_faces}, the validity of \name{}\textsubscript{gen}~\textit{clustered} degrades substantially, confirming that the bottleneck lies more in the rebuilding step rather than in the field prediction.
This highlights the inherent fragility of CAD solids---a single missing or misaligned face barely affects aggregate F1 but is sufficient to violate the closed-shell constraint.
Fig.~\ref{fig:validity_gap_examples} illustrates this concretely: \name{}\textsubscript{gen} and \name{}\textsubscript{recon} produce nearly identical segmentations for the same input, yet a small localized perturbation in the generative sample prevents the rebuilder from closing the shell.

\paragraph{Merits of the generative formulation.}
Despite the additional stochasticity introduced by the ODE solver, the generative variant of \name offers distinct practical advantages that a deterministic encoder cannot provide.
First, the \emph{generative} nature of \name{}\textsubscript{gen} naturally enables drawing multiple samples for the same input and selecting the best one---a strategy already shown to boost validity in prior B-rep generation work~\cite{HolaBRep25,brepgen24}; the results reported in Table 1 in the main paper correspond to only a \emph{single} sample per input and therefore represent a lower bound on achievable quality.
Second, the generative formulation opens up diverse conditioning workflows that are inaccessible to deterministic reconstruction, such as image-to-B-rep and sketch-to-B-rep generation.
Third, recent mesh generation methods~\cite{trellis24,hunyuan,craftsman24,xcube24} have demonstrated that latent generative models scale favorably with data and compute; extending this scaling paradigm to B-rep generation is a promising direction that the \name{}\textsubscript{gen} architecture is well positioned to explore.

\paragraph{Summary.}
Our analysis reveals that \name{}\textsubscript{gen}'s stochastic nature makes it more sensitive to small perturbations in the predicted fields, which are in turn amplified by the current rebuilder's zero-tolerance rebuilding into disproportionate validity failures.
Given the practical merits of the generative formulation---multi-sample selection, diverse conditioning workflows, and favorable scaling properties---improving the robustness of the rebuilder (e.g., through more tolerant stitching heuristics or iterative geometric refinement) emerges as the primary bottleneck and a promising direction for future work.
Scaling up \name{}\textsubscript{gen} with larger datasets and compute, following the trajectory of recent mesh generation methods~\cite{trellis24,hunyuan,xcube24}, is another complementary path toward closing the remaining validity gap.
}

\section{Additional Qualitative Results}

We provide extended qualitative comparisons with baseline methods across diverse CAD models.

\begin{figure*}[t]
    \centering
    \includegraphics[width=0.95\linewidth]{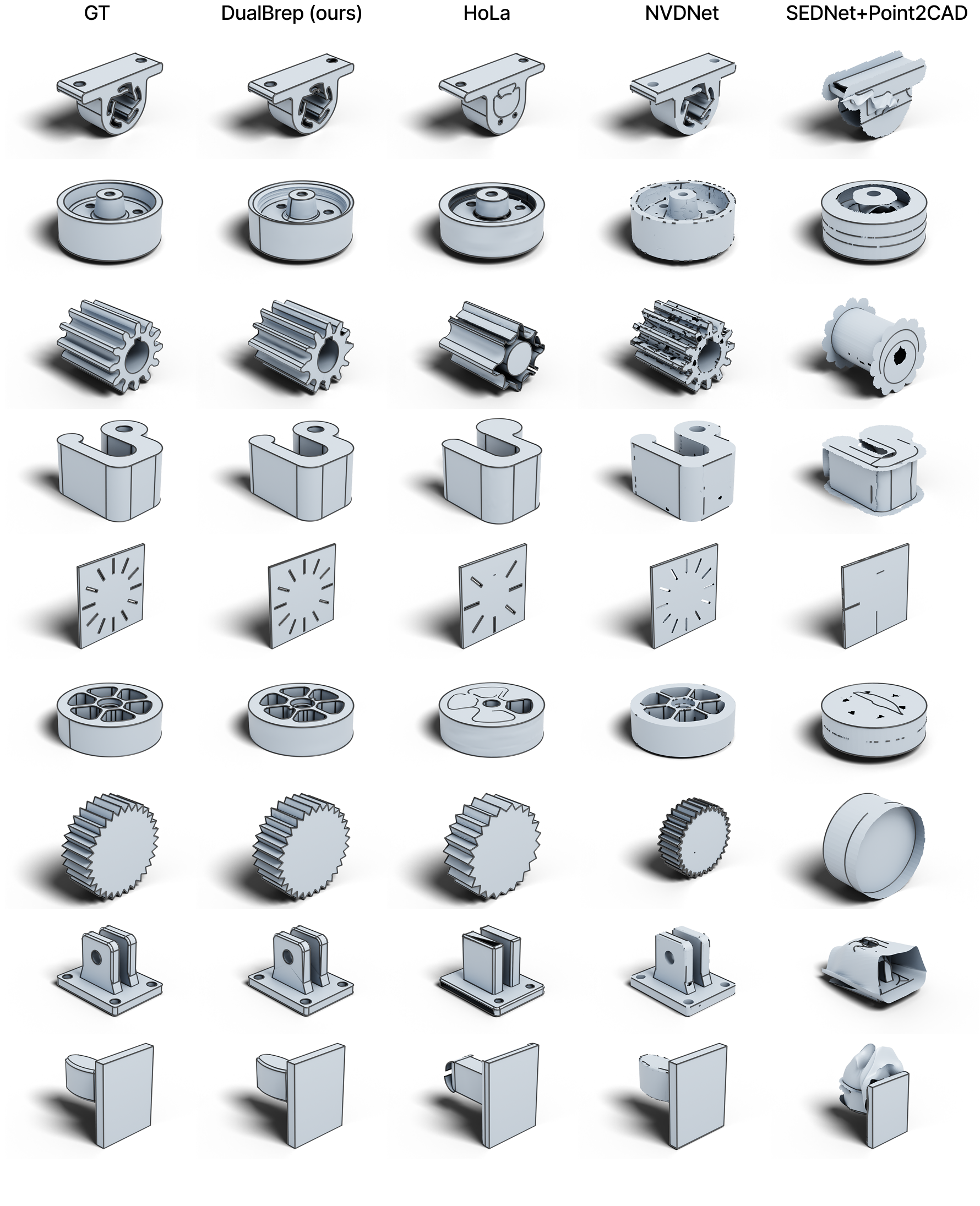}
    \caption{
        Additional comparison on deterministic reverse engineering from \name{}\textsubscript{recon}.
    }
    \label{fig:supp_comparison1}
\end{figure*}

\begin{figure*}[t]
    \centering
    \includegraphics[width=0.95\linewidth]{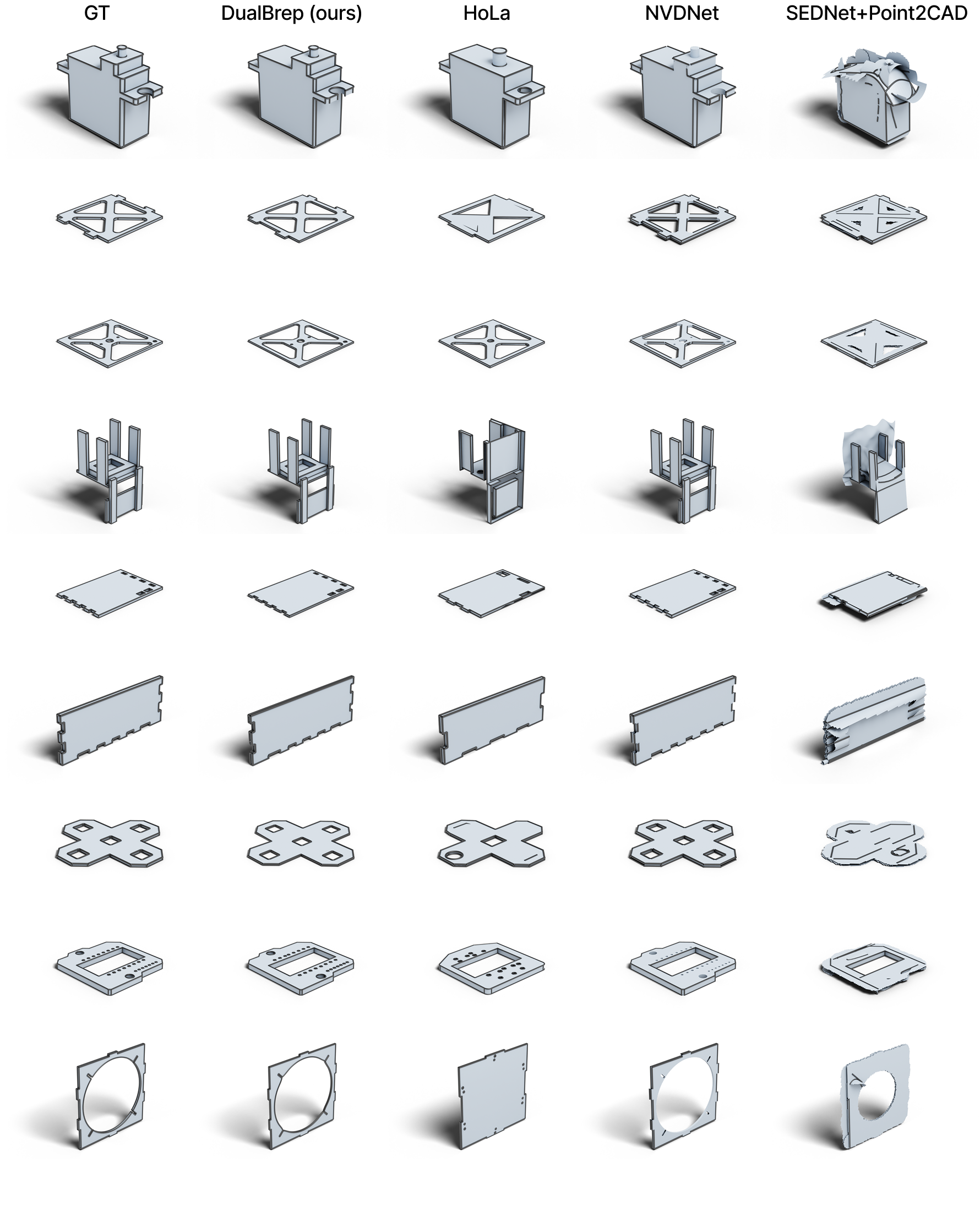}
    \caption{
        Additional comparison on deterministic reverse engineering from \name{}\textsubscript{recon}.
    }
    \label{fig:supp_comparison2}
\end{figure*}

\begin{figure*}[t]
    \centering
    \includegraphics[width=1\linewidth]{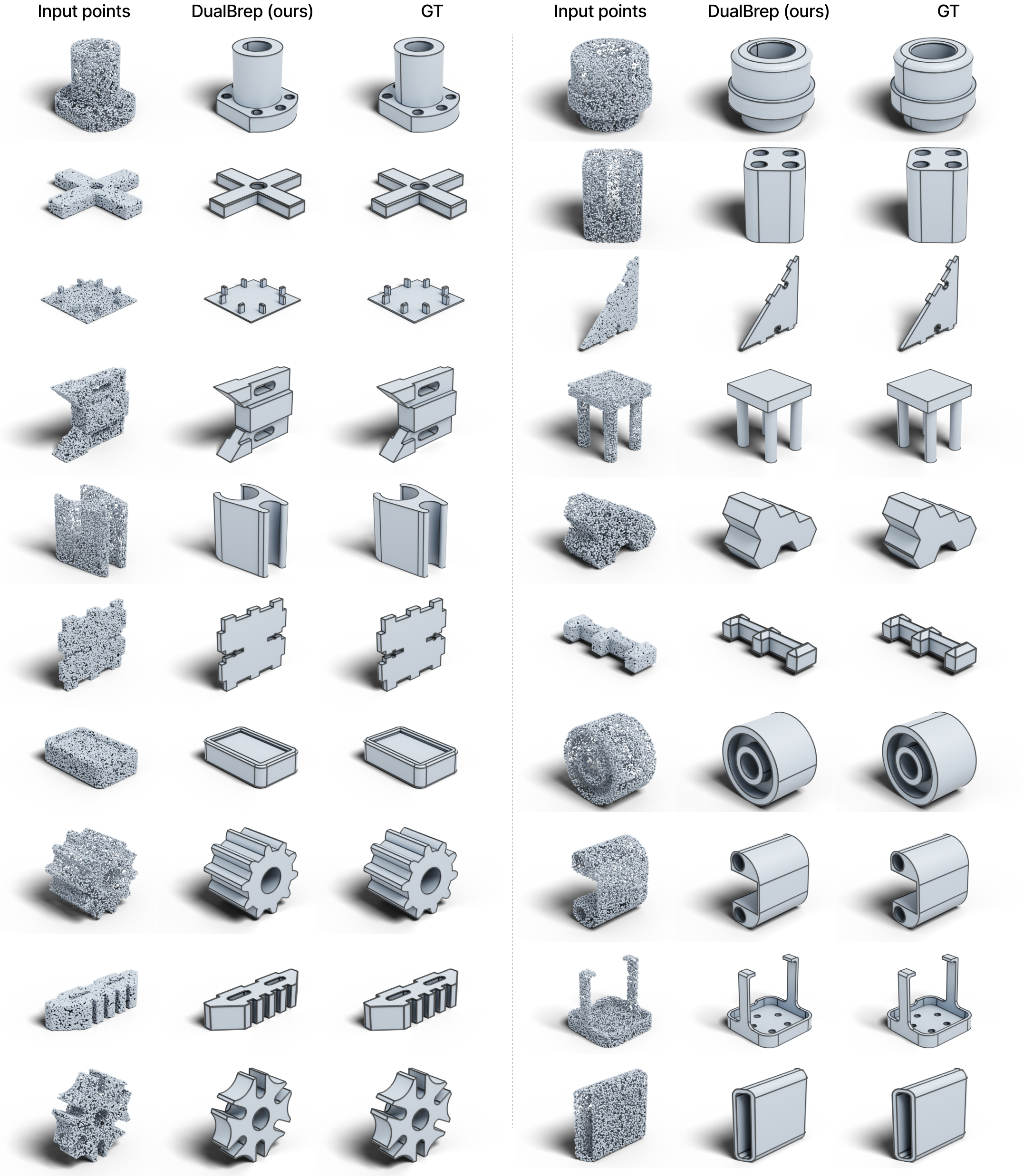}
    \caption{
        Additional reconstruction result on deterministic reverse engineering from \name{}\textsubscript{recon}.
    }
    \label{fig:supp_gallery1}
\end{figure*}

\begin{figure*}[t]
    \centering
    \includegraphics[width=1\linewidth]{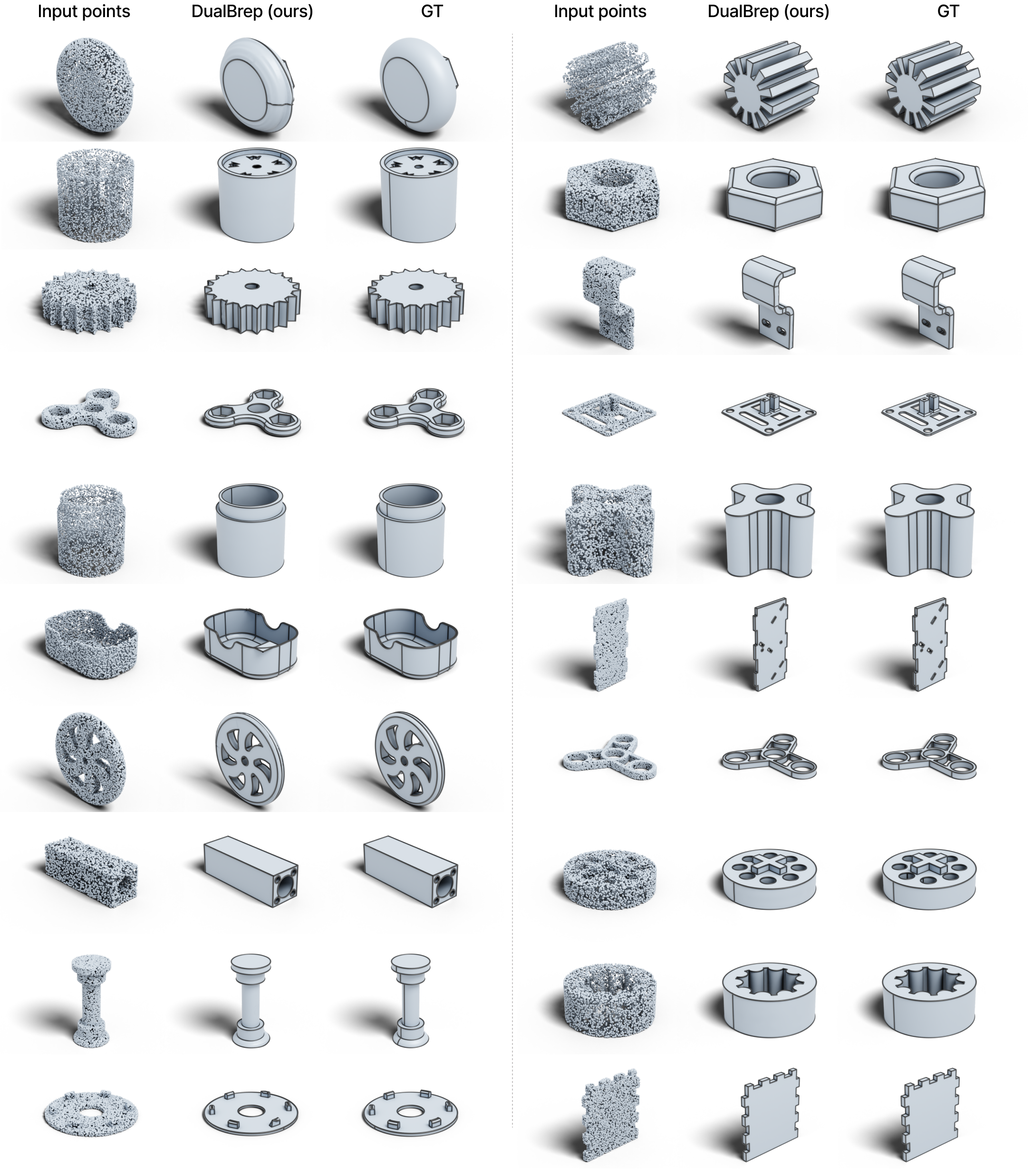}
    \caption{
        Additional reconstruction result on deterministic reverse engineering from \name{}\textsubscript{recon}.
    }
    \label{fig:supp_gallery2}
\end{figure*}

\begin{figure*}[t]
    \centering
    \includegraphics[width=1\linewidth]{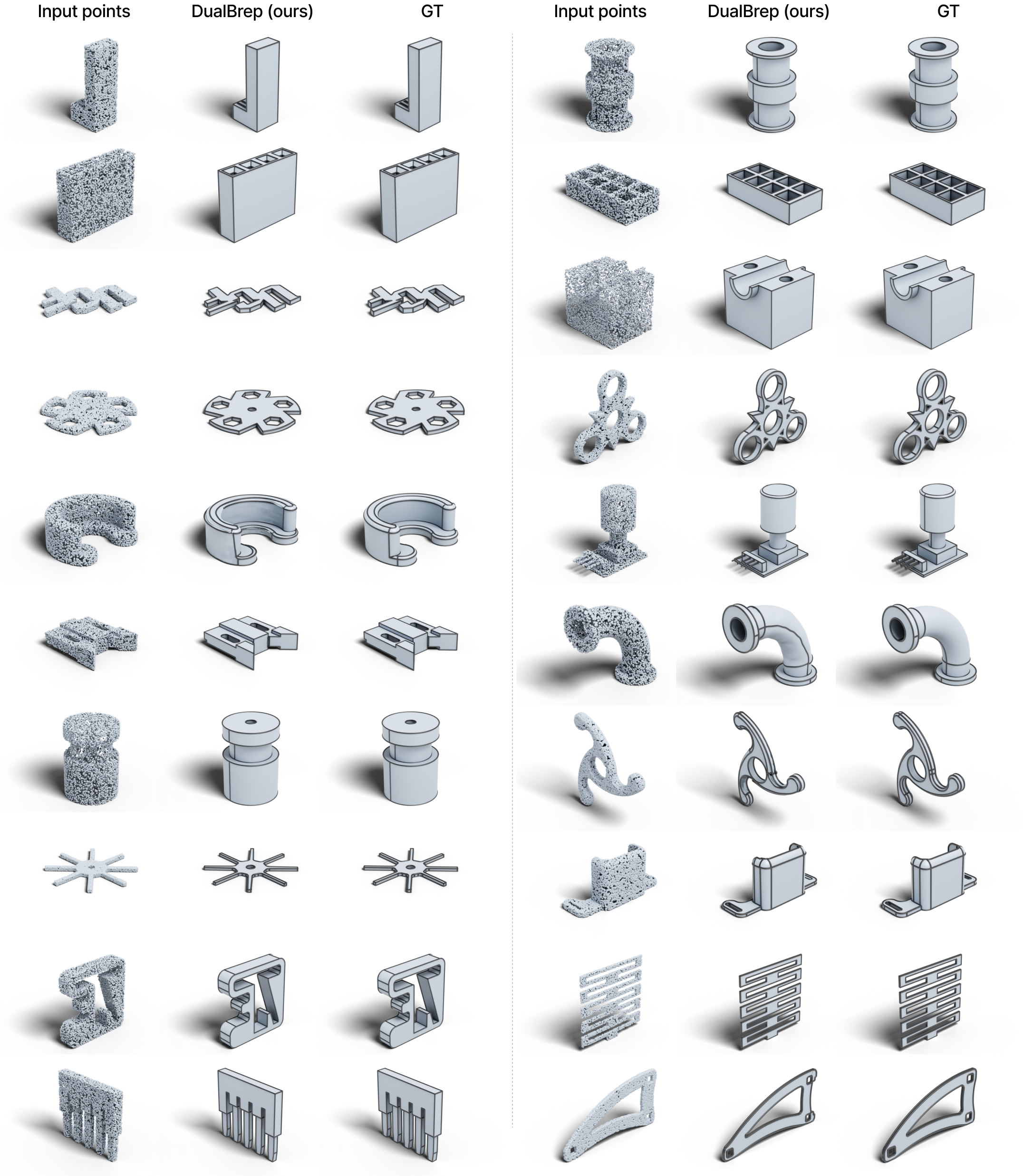}
    \caption{
        Additional reconstruction result on deterministic reverse engineering from \name{}\textsubscript{recon}.
    }
    \label{fig:supp_gallery3}
\end{figure*}

\begin{figure*}[t]
    \centering
    \includegraphics[width=0.9\linewidth]{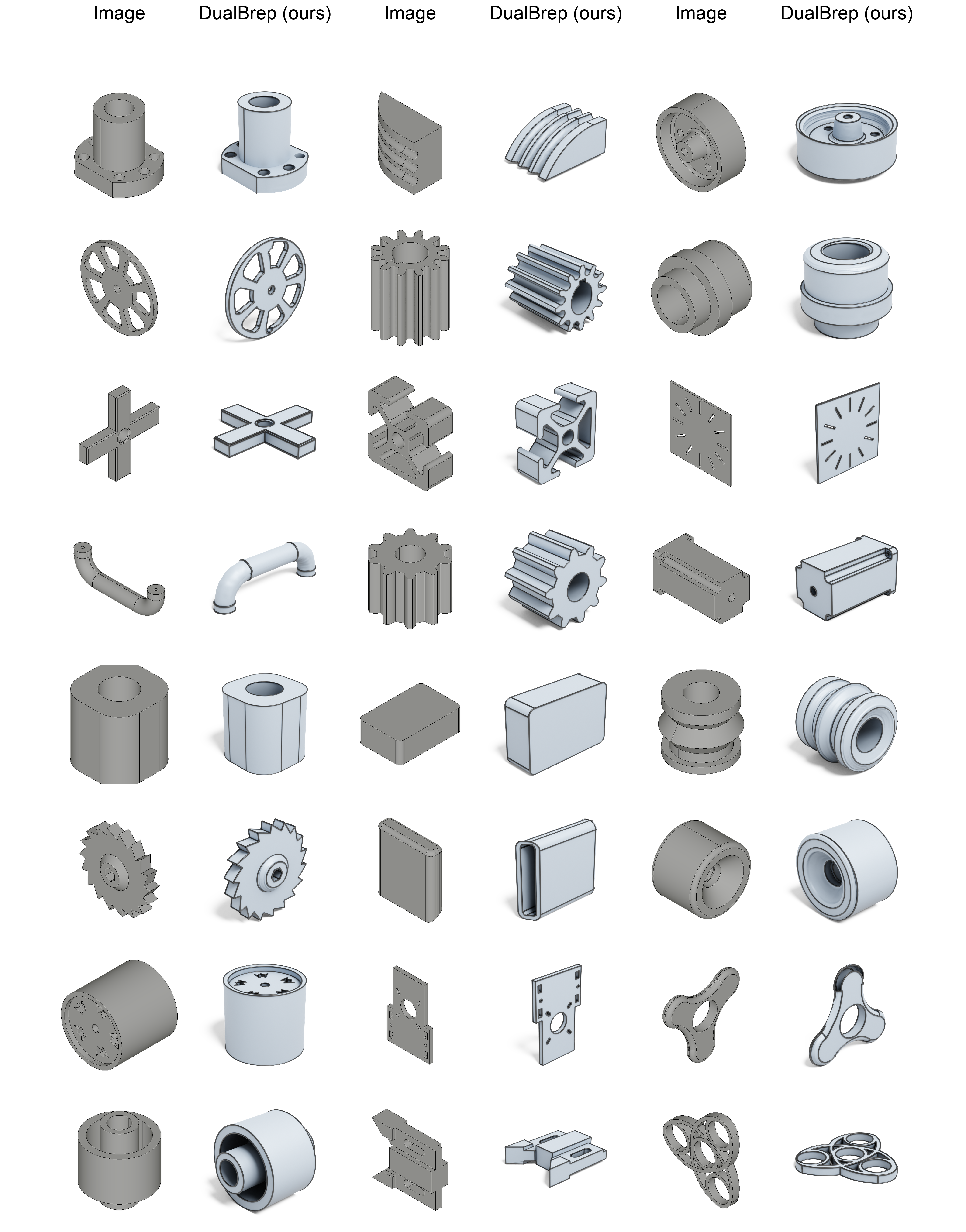}
    \caption{
        Additional conditional generation results from single view images from \name{}\textsubscript{gen}.
    }
    \label{fig:supp_img1}
\end{figure*}

\begin{figure*}[t]
    \centering
    \includegraphics[width=0.9\linewidth]{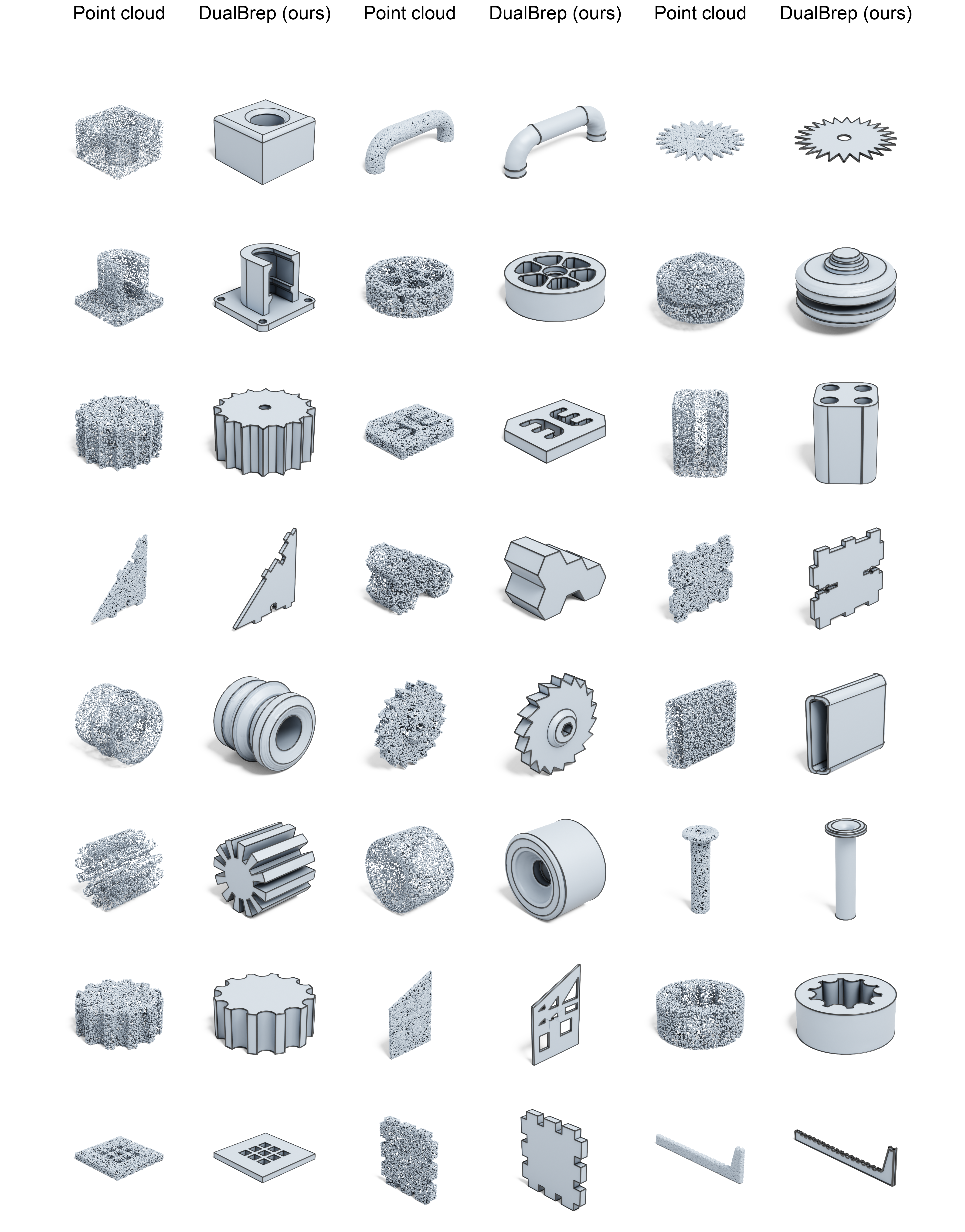}
    \caption{
        Additional conditional generation results from point clouds from \name{}\textsubscript{gen}.
    }
    \label{fig:supp_pc3}
\end{figure*}

\end{document}